# Sources of Emittance in RF Photocathode Injectors:
# Intrinsic emittance, space charge forces due to non-uniformities, RF and solenoid effects

David H. Dowell
SLAC National Accelerator Laboratory, Menlo Park, CA 94025, USA

**Abstract**

Advances in electron beam technology have been central to creating the current generation of x-ray free electron lasers and ultra-fast electron microscopes. These once exotic devices have become essential tools for basic research and applied science. One important beam technology for both is the electron source which, for many of these instruments, is the photocathode RF gun. The invention of the photocathode gun and the concepts of emittance compensation and beam matching in the presence of space charge and RF forces have made these high-quality beams possible. Achieving even brighter beams requires a taking a finer resolution view of the electron dynamics near the cathode during photoemission and the initial acceleration of the beam. In addition, the high brightness beam is more sensitive to degradation by the optical aberrations of the gun's RF and magnetic lenses. This paper discusses these topics including the beam properties due to fundamental photoemission physics, space charge effects close to the cathode, and optical distortions introduced by the RF and solenoid fields. Analytic relations for these phenomena are derived and compared with numerical simulations.

**Introduction**

This paper explores the sources of emittance in the photocathode gun and solenoid system currently used in high brightness injectors. This will be done using a combined analytic and numerical analysis approach to isolate and understand the various mechanisms which generate emittance. Sources of the intrinsic emittance of the cathode, the space charge driven emittance growth near the cathode, emittance due to the gun RF and the optical aberrations of the emittance compensation solenoid are described and compared.

The present work begins with a brief introduction to photocathode injector design philosophy. Then there is a discussion about the connection between the intrinsic emittance and the quantum efficiency. The concept of using the tensor properties of the electron's effective mass to reduce the intrinsic emittance while maintaining good QE is explained. Next the emittance due to transverse space charge forces produced by non-uniform emission is derived using an analytic model with some mathematical approximations. Good agreement with experimental results indicates this model provides a useful explanation of the underlying physics despite its simple result. The first- and second-order emittances produced by the time-dependence of the gun's RF fields is described. While this effect is absent in the DC gun, a modified version of the formula is still useful for computing the RF emittance of the first accelerator section after the gun. The discussion then turns to the extensive topics of optical distortions and aberrations. The solenoid's chromatic, geometric, and anomalous quadrupole field effects are described and analytic expressions for their emittances are derived. The paper concludes with a summary comparing these phenomena.

**Generic Photocathode Injectors with DC and RF Guns**

The basic architecture of the photocathode injector is mostly determined by the cathode field and the gun voltage. The injector using a low cathode field gun typically needs a longer gun-to-linac section with two solenoids and a buncher cavity as shown in Figure 1a. Because it uses a combination of a DC gun and a RF linac this injector is referred to as the *DC/RF injector*. If the cathode field is high and the energy out of the gun is relativistic then this injector type is the *pulsed RF injector*. In this case, the extreme acceleration from the cathode naturally produces an electron bunch with the desired length for injection into the linac. Thus, only one solenoid is needed to compensate the emittance and match the beam into the linac as shown in Figure 1b.

In the DC/RF injector, the low cathode field requires longer bunches to preserve the emittance. These longer bunches are then compressed using a RF buncher cavity and ballistic compression in a drift before injection into the first RF accelerator section. In addition to compressing the bunch, the injector's gun-to-linac region needs to match the beam into the linac while minimizing any emittance growth. The emittance aberrations are controlled using two solenoids. The first solenoid produces a small beam through the RF buncher. To velocity-bunch the beam before it enters the linac, the buncher-beam phase is near a zero crossing to give the bunch a large energy chirp. Besides compressing the bunch, this large energy chirp generates chromatic emittance in the second solenoid. As described later in this paper, both the RF emittance of the buncher and the chromatic emittance of the solenoid scale as the square of the transverse beam size. Therefore, as a compromise, the beam waist is placed downstream of the RF buncher but before Solenoid 2, to make the beam small at both locations. This injector configuration has successfully generated high-peak brightness beams at record high average current [1][2].

The DC/RF gun system uses a pulsed drive laser to produce the electron bunches needed for injection into the RF linac. Since the gun is DC, it can produce beam with any frequency or pulse format desired. This gives the HVDC gun a distinct advantage over the RF gun. However, the photocathode injector using a HVDC gun operates at a lower cathode field and gun voltage. Therefore, the electron bunches emerging from the gun needs to be longer and larger to mitigate space charge forces. [3]

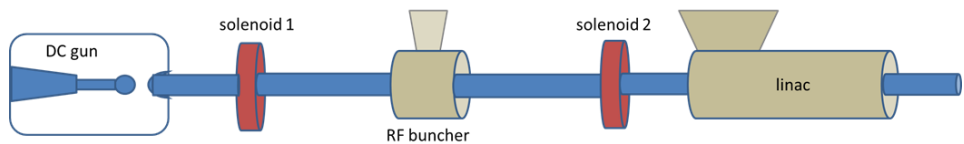

Figure 1a(color): The standard DC/RF injector configuration. This layout is also used with a low-frequency RF gun instead of the DC gun. Both the DC and low-frequency (UHF, e.g., the 187 MHz gun at LBNL) guns use low cathode fields and gun voltages to operate at CW duty factor and high average current.

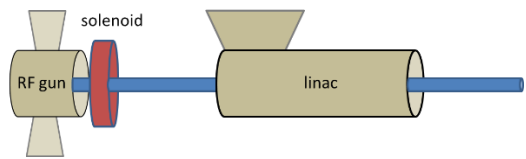

Figure 1b(color): The standard injector configuration for high field, pulsed RF guns.

Pulsed RF guns can achieve high cathode fields which greatly mitigate the space-charge forces by rapidly accelerating the photoelectrons to relativistic energies. The high-field configuration commonly used with pulsed RF guns is shown in Figure 1b. In this configuration, a single solenoid is used to produce a beam waist at or near the linac's entrance. As shown by emittance compensation theory [4], the distance between the gun and the linac entrance giving the lowest injector emittance is a multiple of one-quarter wave at the bunch's plasma frequency. Theory and experiment show the bunch radius and divergence oscillate along the beamline between the gun and the linac, with these parameters repeating themselves every ¼ of the bunch's plasma wavelength.

The time-dependence of the higher frequency RF field can be used to chirp the bunch energy and control the bunch length out of the gun. The slice-to-slice energy chirp along the bunch can then be arranged to maintain the laser pulse or even compress the bunch. Hence, the beam out of the high-field gun requires no compression or further acceleration before injection into the first linac. However, it does need to be "emittance matched" into the first accelerator linac. This is done using a single magnetic solenoid to both cancel the gun's large RF defocusing strength, and to compensate for linear space-charge forces on the transverse phase space as shown in Figure 1b. This injector configuration has produced record peak brightness beams and is arguably the current state-of-the-art in pulsed RF injectors for x-ray free electron lasers [5].

The pulsed photocathode RF gun consists of n+1/2 cells where each full cell is $\lambda_{rf}/2$ long with the cathode at a wall in the middle of the ½ cell which is then $\lambda_{rf}/4$ long. Guns have been built and operated with n ranging from 0 to 4. Optimizing with beam simulation codes has determined that beam performance is improved if the half-cell is slightly longer at 0.3 $\lambda_{rf}$ rather than 0.25 $\lambda_{rf}$. Hence most high field RF guns are thus (n+0.6)$\lambda_{rf}/2$ long. The standing wave RF guns have been demonstrated at frequencies from 144 MHz to 17 GHz. In general, the higher RF frequencies (~GHz and higher) can operate with high peak cathode fields (>40 MV/m) to rapidly accelerate the beam to relativistic energy and mitigate space charge forces. However, the high field comes at the expense of duty factor which is a fraction of a percent at s-band (~3 GHz) and higher frequencies. Lower RF frequency guns are capable of CW operation albeit by limiting the peak cathode field. Further descriptions of RF guns both normal conducting and superconducting can be found in Chapters 1 and 3 of Ref [3].

Figure 1b shows a magnetic solenoid near the high-field RF gun exit. This focusing solenoid cancels the strong RF defocusing of the beam by the gun exit field. This solenoid also matches the bunch to the first accelerator section or linac for optimal emittance compensation [6]. In addition, there is often another coil (not shown in Fig. 1b) positioned just behind the cathode for zeroing or bucking the gun solenoid's fringe field at the cathode. If the cathode magnetic field is not zero, the electrons acquire canonical angular momentum and thus emittance. In this paper, the initial angular momentum is assumed to be zero.

In both DC and RF guns, the electron bunches are produced from a photocathode with a drive laser phase-locked to the RF master oscillator. The type of laser used depends upon the cathode material and the duty factor of the system. The cathode material determines the laser wavelength and pulse energy needed given the cathode quantum efficiency (*QE*) and wavelength sensitivity. In addition, the desired charge and intrinsic emittance are also important factors to consider when designing an injector system. While the largest uncertainly still lies in the cathode properties of QE and intrinsic emittance [7], there has been considerable progress in understanding the physics and practical aspects of cathodes as documented in the Photocathode Physics for Photoinjectors workshops held only on even-numbered years since 2010 [8].

**Evolution of the emittance from the cathode through the injector**

As the beam is born and accelerated from the cathode it undergoes processes and forces which interact with it and add to its emittance. Figure 2 attempts to make sense of these complex interactions by spatially ordering these processes as a function of distance from the cathode. The flow chart indicates there are five distance scales (yellow-boxes) over which the beam experiences emittance generation and growth. The physical properties or characteristics (grey-boxes) are combined as inputs to 'and-gates' which generate emittance (green-boxes) and more properties/characteristics (grey-boxes). These properties can then combine with another set of external properties like non-linear focusing to produce yet more emittance and more properties as the beam propagates down the beamline. As will be shown this spatial flow of the emittance growth and its interactions provides a useful basis for analyzing the sources of emittance in the photocathode injector.

The Figure 2 chart shows the emittance and properties/characteristics flowing from left to right which interact in a series of 'and-gates'. The chart begins with the <Cathode Material Properties> and <Applied Field> interacting in the first 'and-gate' to generated the <Intrinsic Emittance>. Adding with <Surface Roughness> in the next 'and-gate' gives the <Rough Surface Emittance>. On the other hand, the intrinsic emittance isn't necessary to generate the <Applied Field Emittance> some tens of microns from the surface. The emission processes during the laser pulse both below and at the surface establishes/determines the <Cathode Emission Properties> such as response time, linearity, uniformity and image-charge-bunch interactions which are most influential at distances of microns to millimeters from the surface where and while the bunch is still emerging from the cathode.

At millimeters from the surface, the <Cathode Emission Properties> are 'added' in the fourth gate with the <Transverse Density Modulation due to the Rough Surface> and the drive laser's <3D Laser Shape> to produce the <6D Phase Space> distribution and two more emittances. At this location, typically a few 10's of mm from the cathode, the electron bunch is fully formed with the bunch tail separate from the cathode surface. <6D Phase Space> then 'adds' with <Non-Linear Focusing and Alignment Errors> for use in relativistic transport codes with space charge to obtain the <Emittances due to Optical Aberrations, Space-Charge and other effects> during acceleration and compression of the electron bunch.

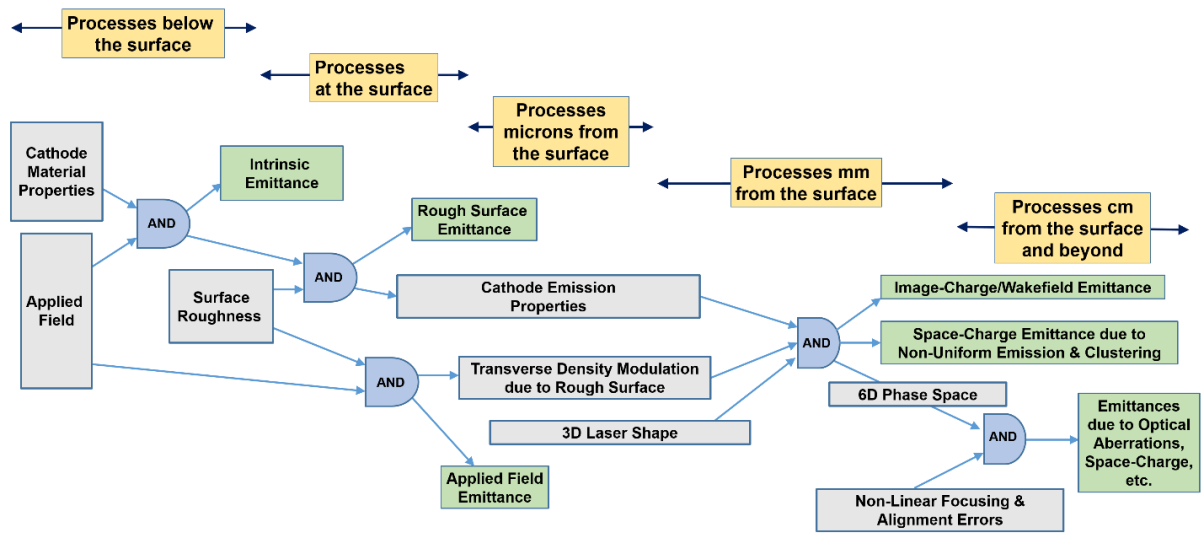

Figure 2(color): Flow chart illustrating the various emittance-generating processes vs. distance from the cathode surface.

## Photoemission Theory for Metal Cathodes

In this section the quantum efficiency and intrinsic emittance of a metallic photocathode are derived using the Spicer three-step model [9]. In this model photoemission is separated into the steps of photon absorption, electron transport to the surface and electron escape into the vacuum. The discussion begins with a brief description of the electric potential which binds the electrons in the cathode and the potential barrier over which they must pass to escape into the vacuum. Next it is shown how these work functions are used in expressions for the quantum efficiency and intrinsic emittance in terms of the electron excess energy and the electron's effective mass.

### *The electrical potentials at the metal-vacuum interface*

The forces on an electron near the cathode surface are due to three electric potentials: 1) the material work function, $\phi_W$, produced by a thin layer of electrons forming a surface dipole layer at the cathode-vacuum boundary [10], 2) the image potential due to the electron's equal and opposite image in the metallic surface and 3) the external field which in this case is the RF field. These potentials are plotted in Figure 2. The combined, external fields produce a potential barrier outside the surface and with a height the Schottky work function below the vacuum energy. The quantum efficiency and the intrinsic emittance using these potentials along with the three-step model of photoemission have been derived elsewhere [11]. The reformulated results which now include the effective mass are given here.

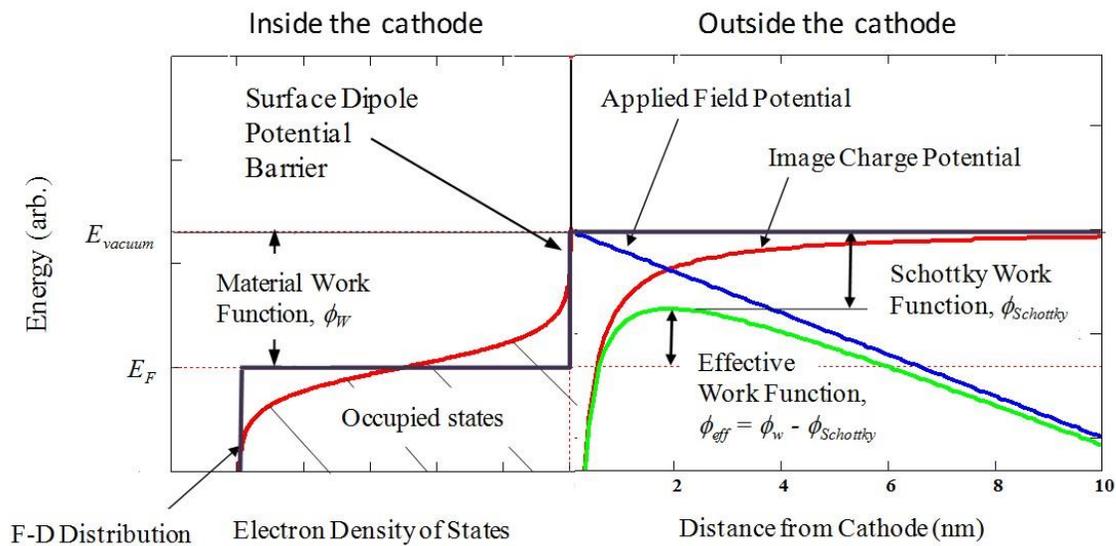

Figure 3(color): The occupied electron energy levels or electron density of states inside the cathode (left) and the electric potentials (right) at the cathode-vacuum interface. The distribution of occupied states is given by the Fermi-Dirac (F-D) function, $f_{FD}(E)$ (solid red line, inside cathode). For a metal at 300 degK, $f_{FD}(E)$ can be replaced the Heaviside step function with its step at $E_F$, indicated by the heavy solid line. Outside the cathode there is the potential due to the image charge of the electron (red) as well as the applied field potential (blue). The sum of the image and applied potentials (green) forms a potential barrier which reduces the material work function by the Schottky work function.

Inside the cathode the fermionic electrons fill energy states in pairs up to the Fermi energy, $E_F$, which positions the material's work function, $\phi_W$, below the vacuum state energy, $E_{vacuum}$. In this model the electron energy distribution is given by the Fermi-Dirac function which in turn is replaced by the Heaviside step-function to simplify the calculation. The step-function is a very good approximation at ambient temperatures. In the absence of all other forces the material work function is defined as the energy an electron needs to escape the cathode material. Outside the cathode, the electron's image charge and an accelerating applied field combine to form a shallow potential barrier approximately a few nm from the cathode surface, depending upon the strength of the applied field. The barrier height is a Schottky work function, $\phi_{Schottky}$, below the vacuum state energy which reduces the material work function to give the effective work function,

$$\phi_{eff} = \phi_W - \phi_{Schottky} \tag{1}$$

The photoemission process assumes there is no quantum mechanical tunneling therefore electrons require energies greater than the barrier height of $\phi_{eff}$ to escape. Thus, when excited by photons having energy $\hbar\omega$, electrons in occupied states with energies between $E_F - (\hbar\omega - \phi_{eff})$ to $E_F$ can escape into the vacuum. After emission, electrons can have energies from zero to $\hbar\omega - \phi_{eff}$. Using a step-function for the energy distribution of occupied states inside the cathode, the emitted electron energy spectrum has a full width of $\hbar\omega - \phi_{eff}$. It is this energy spread which causes the intrinsic emittance and the yield of this energy spectrum determines the *QE*. Due to its relevance to both the *QE* and intrinsic emittance discussed next, it is useful to define the excess energy, $E_{excess}$ as

$$E_{excess} \equiv \hbar\omega - \phi_{eff} \tag{2}$$

*Quantum efficiency and intrinsic emittance theory*

The quantum efficiency and the photoelectric emittance for a cathode can be derived by following the assumptions of Spicer's model of photoemission [1]. This model defines the following three steps: 1) absorption of photon by a bound electron, 2) excited electron travels to the surface, and 3) electron escapes to vacuum. The mathematical representation for these steps is given in Eqn. (3) The limits of the energy integral reflect using a step-function for the initial occupied energy state distribution. The limits for the $\theta$-integral are from maximum escape angle to normal incidence. The maximum escape angle is discussed later. The azimuth angle integration limits assume the photon excited electrons motion is isotropic inside the cathode. Details of evaluating the electron-electron scattering length and performing these integrals are given in Ref. [11].

$$QE = (1-R(\omega))\frac{\left[\int_{E_F+\phi_{eff}-\hbar\omega}^{\infty} dE\, N(E+\hbar\omega)(1-f_{FD}(E+\hbar\omega))N(E)f_{FD}\right]\left[\int_1^{\cos\theta_{max}(E)} d(\cos\theta) F_{e-e}(E,\omega,\theta)\int_0^{2\pi} d\varphi\right]}{\left[\int_{E_F}^{\infty} dE\, N(E+\hbar\omega)(1-f_{FD}(E+\hbar\omega))N(E)f_{FD}\right]\left[\int_1^0 d(\cos\theta)\int_0^{2\pi} d\varphi\right]} \tag{3}$$

Since the *QE* is defined as the number of emitted electrons per incident photon, Step 1 simply involves the fraction of incident photons which are absorbed, *1-R(ω)*. The reflectivity, *R(ω)*, is obtained from the Fresnel optical relations using the complex index of refraction. The optical absorption depth, $\lambda_{opt}$, used in the second step is given by the imaginary part of the index of refraction also using these optical relations. At 253 nm the reflectivity for copper at normal incidence is approximately 0.3, making the Step 1 factor 0.6.

The Step 2 factor is given by the second square bracket and gives the fraction of excited electrons which arrive from below the surface. Here the important parameters are the optical absorption depth (described above as $\lambda_{opt}$) and the energy-averaged electron mean free path between scattering events ($\bar{\lambda}_{e-e}$). The excited electrons can scatter either with the lattice via electron-phonon scattering or with the valence electrons. For a good metal, such as copper, electron-electron scattering dominates while electron-phonon scattering is important for semi-conductor cathodes. For copper illuminated at normal incidence with 4.86 eV photons the optical absorption depth is approximately 10 angstroms and the electron-electron scattering length for energies near the Fermi level is approximately 30 angstroms. Using these values in the bracket 2 term indicates the fraction of excited electrons reaching the surface is approximately 0.2.

The term for Step 3 involves integrals over the electrons' energy spectrum and the polar, $\theta$, and azimuth, $\Phi$, angles the electrons have with respect to the surface normal. The energy integration limits of the numerator correspond to the energy range needed to escape over the potential barrier. The energy limits of the integral in the denominator correspond to all the electrons the photon can excite, that is, down to the photon energy below the Fermi level. In passing, it important to point out that other functions for the density of states can and should be used for other cathode materials or at higher photon energies reaching further below the Fermi level. Step 3 also assumes the excited electrons inside the cathode have an isotropic angular distribution. This means the photon's momentum is not conserved in the 3-step model, the electron has no knowledge of the photon's initial direction. However, since the transition is direct, the energy is conserved. The $\theta$-integration limits are determined by the continuity of the transverse momentum across the cathode-vacuum boundary. It can be shown that the maximum polar angle for which an electron with an initial energy $E$ can escape is $\cos\theta_{max} = \sqrt{\frac{m}{m^*}\left(\frac{E+\hbar\omega-E_F-\phi_{eff}}{E+\hbar\omega}\right)}$. The fraction of electrons in occupied states which have enough energy and are within the angular escape cone is approximately 0.04, and the fraction of electrons within the maximum internal escape angle is approximately 0.01, for $m^* = m$.

Therefore, the $QE$ is low for metals because first the photon's energy can reach only a limited number of occupied electronic states, and second there is a small acceptance angle at the surface into which the electrons can escape. Reducing reflectivity to zero increases the $QE$ about a factor of two while eliminating e-e scattering would result in approximately five-times the $QE$. In other words, for photon energies less than a volt greater than the effective work function, the emission yield is only a few percent of the total number of energetically available electrons. And due to refraction at the surface, only electrons within an internal angle of incidence less than ~10 degrees can escape. For copper this is only one percent of the excited electrons.

Performing the integrations in Eqn. (3) gives the quantum efficiency with the effective mass,

$$QE = \frac{1-R(\omega)}{1+\frac{\lambda_{opt}}{\lambda_{e-e}}} \frac{E_F+\hbar\omega}{2\hbar\omega}\left(1-\sqrt{\frac{m}{m^*}\left(\frac{E_F+\phi_{eff}}{E_F+\hbar\omega}\right)}\right)^2 \quad (4)$$

Similarly, the 3-step model can be used to compute the variance of the transverse momentum giving the normalized intrinsic emittance for a transverse rms beam size, $\sigma_x$, in terms of the excess energy [12] and the effective mass [13] as,

$$\epsilon_{intrinsic} = \sigma_x\sqrt{\left(\frac{m^*}{m}\right)\left(\frac{\hbar\omega-\phi_{eff}}{3mc^2}\right)} \quad (5)$$

These results show the $QE$ and intrinsic emittance are both increasing functions of the excess energy as commonly accepted. But equally important is the $\sqrt{m^*}$-dependence which could allow achieving ultra-low intrinsic emittance from a practical cathode a real possibility. The effective mass and the general cathode material properties needed to obtain low intrinsic emittance are discussed next.

**Effective Mass Effects on the QE and Intrinsic Emittance**

Eqns. (4) and (5) show that in addition to the excess energy, the emittance and QE also depend upon the effective mass the electron has before emission. As pointed out by Berger *et al.* [13], the intrinsic emittance is proportional to $\sqrt{m^*}$. Therefore, the effective mass should be as small as possible to give a very small transverse momentum and thus an ultra-low intrinsic emittance. Eqn. (5) shows the QE follows the opposite trend by growing with increasing $m^*$. This is because the larger effective mass reduces $\cos\theta_{max}$ (see above equation in text) and thereby increases the internal escape cone angle. Thus, a small effective mass leads to low QE. And it appears there is no easy solution since low intrinsic emittance requires $m^*/m \ll 1$, yet high QE needs $m^*/m \gg 1$. This is the same situation as in the case of near threshold photoemission, where decreasing the excess energy reduces the intrinsic emittance but it also lowers the QE [14]. However, there is a possible path around this apparent law of nature.

The effective mass is a tensor quantity and for some materials it can have very different values for components along orthogonal axes. In addition, it's important to note that the emittance is driven by the electron's transverse dynamics while the QE results from its longitudinal motion. Therefore, an anisotropic structured, crystalline cathode could have internal electrons with large and small effective masses in orthogonal directions. Then by orienting the axis with large effective mass normal to the surface (along the electron's longitudinal direction), which naturally places the small effective mass axis along the transverse direction. With this arrangement, this small transverse effective mass will produce a low intrinsic emittance, while the large longitudinal effective mass will give preserve the QE.

**Space Charge Emittance near the Cathode due to Non-Uniform Emission**

Extensive experimental and theoretical studies have been performed to understand the effect of non-uniform emission upon beam quality in space charge dominated beams. See for example Refs. [15,16]. This work established the transverse uniformity specifications for low emittance beams of sufficient quality to drive x-ray FELs. The influence emittance and other beam characteristics have upon xfel performance were determined from simulations and analytic theories. Recent xfel experiments performed at the SLAC Linac Coherent Light Source measured how non-uniform emission affects the xfel performance [17]. In these studies, laser patterns consisting of regular rectangular meshes and circular distributions resembling donut, bagel and Airy-like patterns were imaged onto the cathode and the emittance and xfel output and gain were measured. A space charge model was developed to analyze these data. For the rectangular mesh patterns used in the experiment the model is in good agreement with emittance measurements. In this section this space charge model will be discussed and emittance will be given in terms of the number of spatial modulations across the cathode diameter and the transverse variation in peak current.

The space charge beamlet model analyzes the regular rectangular mesh pattern to derive the emittance growth due to regular high-spatial frequency (several cycles across the beam diameter) patterns. A beginning assumption is that immediately after emission the space charge forces can be computed classically. Then due to the non-uniform emission the electrons within and at the edges of the beamlets will feel a radial space charge acceleration and the beamlets will expand. When the beamlets overlap, on average the beam becomes more uniform and the space charge force diminishes and the electrons continue to expand with a constant radial velocity. The transverse emittance results from this radial velocity.

In most RF guns the cathode field is high and the beamlets overlap a few tens of picoseconds after emission. At the time of overlap, the beam is not yet relativistic which justifies using classical electrostatics in the derivation. This point is discussed in more detail below. Once the beamlets merge the emittance stops growing due to the nearly uniform density distribution and the onset of relativistic effects. This approach can also be used to compute the emittance of other patterns such as the donut and bagel (for example, as in Ref. [17]), since the physical assumptions can be applied these any emission pattern. In this section the regular rectangular pattern is analyzed to develop an expression useful in the spectral analysis of the high spatial frequency variations in the emission. The extension of the theory into a general Fourier analysis of the spatial distribution will be left for future studies.

An alternative approach to the theoretical analysis of non-uniform assumes the emittance results from the beam's free energy defined as the potential energy difference between the initial non-stationary and final stationary beam distributions [16]. In the present case of an array of beamlets, the free energy is the transverse kinetic energy and hence produces transverse emittance. This is, in fact, just what's being computed in the expanding beamlet model. It begins with a non-stationary distribution of a regular pattern of beamlets whose potential energy is converted into kinetic energy and emittance as it becomes a stationary or static uniform distribution expanding with constant radial velocity.

The beamlet space charge model assumes a beam transverse distribution with overall radius $R$ and full length $l_b$ composed of many beamlets arranged in a rectangular pattern as shown in Figure 4. Each beamlet has an initial radius $r_0$ with center to center spacing of $4r_0$ in a rectangular grid. The transverse space-charge force causes each beamlet to expand and merge with its neighboring beamlets. This radial acceleration gives the beamlets additional transverse momentum leading to larger emittance for the total beam. A basic assumption of the model is that the transverse space charge force goes to zero once the beamlets merge and form an approximately uniform distribution. Therefore, after merging the non-uniformity space charge emittance becomes constant. The theory developed here indicates the beam is born with a constant emittance and remains so until the beam becomes uniform due to the overlap. At this point, the space-charge forces diminish due to merging beamlets. Simulations and analytic modeling of this geometry show the beamlets overlap within tens of picoseconds, therefore the non-uniformity emittance is generated very close to the cathode before the beam can become relativistic for even the very highest cathode RF fields. It is interesting to note that the electrons are still non-relativistic and the beamlets are merging at the head of each bunch even while the tail electrons are just leaving the cathode.

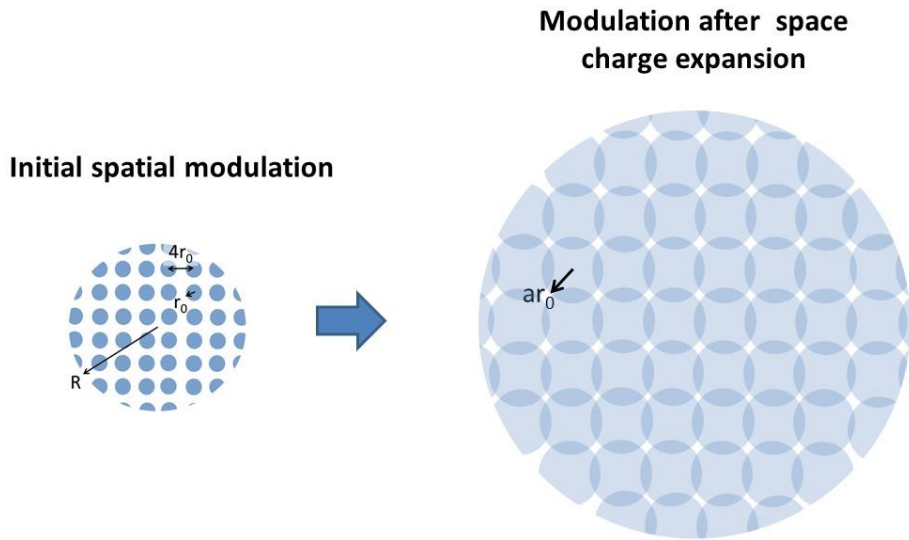

Figure 4(color): Modulation patterns used to compute the space-charge emittance. Left: The initial pattern on the cathode consisting of a rectangular array of circles with radius $r_0$ and a spacing of $4r_0$ within a full beam radius $R$. Right: Schematic view of the beamlet pattern after expansion due to transverse space-charge forces. The integration of the transverse force ends when the beamlets with radius $ar_0$ begin to overlap and form a quasi-uniform distribution.

The derivation for the radial envelope of a beam with uniform charge density during acceleration begins with the equation of motion for an electron at the beam edge. Reiser gives an excellent discussion and justification for the following equation of motion of the boundary of the beam [18],

$$\frac{d^2 r_m}{dz^2} = r_m'' = \frac{K(z)}{r_m} \tag{6}$$

Here K is the relativistic generalized perveance, first defined by Lawson in the late 1950's [19], and is given as

$$K \equiv \frac{I}{I_0} \frac{2}{(\beta\gamma)^3} \tag{7}$$

In this expression, $I$ is the peak current of the beam out to the envelope radius, $r_m$, and $I_0$ is the characteristic current. The characteristic current dependents upon the charge and mass of the beam particle. For electrons, it is given by [18]

$$I_0 \equiv \frac{4\pi\epsilon_0 mc^3}{e} \cong 17000 \; amps \tag{8}$$

The beam is assumed to initially have zero energy spread and position-dependent velocity $\beta(z)$ and energy $\gamma(z) = 1 + \gamma' z$. The model assumes the electrons begin at rest at the cathode ($\gamma = 1, z = 0$) and experience constant acceleration thereafter due to the applied electric field $E_a$. The electron's normalized rate of energy change along the longitudinal axis is defined as

$$\gamma' \equiv \frac{E_a}{mc^2} \tag{9}$$

As expected, this longitudinal acceleration plays a key role in the beam's transverse dynamics.

The radial envelope equation of motion is solved by first multiplying both sides of Eqn. (6) by $r_m'$ so one can write,

$$r_m' r_m'' = K \frac{r_m'}{r_m} \quad \rightarrow \quad \frac{1}{2}\frac{d}{dz} r_m'^2 = K \frac{d}{dz} \ln(r_m) \tag{10}$$

The generalized perveance, $K$, does not depend upon $r_m$. This is because the same charge and hence current is always enclosed by the envelope radius, $r_m$. This property, in fact, is used to define the beamlet and leads to the following integral equation,

$$\int_{r_{m,0}'^2}^{r_m'^2} d\,(r_m'^2) = 2K \int_{r_{m,0}}^{r_m} d\,\ln(r_m) \tag{11}$$

Integrating both sides gives

$$r_m'^2 - r_{m,0}'^2 = 2K \ln\frac{r_m}{r_{m,0}} \tag{12}$$

For now, assume the initial angle is zero, $r'_{m,0} = 0$, then the next integration becomes,

$$\int_{r_{m,0}}^{r_m} \frac{dr_m}{\sqrt{\ln\left(\frac{r_m}{r_{m,0}}\right)}} = 2\sqrt{\frac{I}{I_0}} \int_0^{z_e} \frac{dz}{(\beta\gamma)^{3/4}} \tag{13}$$

The initial, z=0, envelope radius is $r_{m,0}$ and the initial radial angle is $r'_{m,0}$. The upper limit on the z-integral is denoted by $z_e$ for the end of the z-integration. Expressing the left-hand-side in terms of the dimensionless variable $x = r_m/r_{m,0}$ one can write

$$\int_{r_{m,0}}^{r_m} \frac{dr_m}{\sqrt{\ln\left(\frac{r_m}{r_{m,0}}\right)}} = r_{m,0} \int_1^{r_m/r_{m,0}} \frac{dx}{\sqrt{\ln x}} \tag{14}$$

Figure 5 shows that this integral is approximated reasonably well by

$$\int_1^x \frac{dx}{\sqrt{\ln x}} \cong \frac{5}{2}\sqrt{x-1} \qquad \text{for} \quad x \geq 1 \tag{15}$$

The integral on the LHS of Eqn. (13) becomes

$$\frac{5}{2} r_{m,0} \sqrt{\frac{r_m}{r_{m,0}} - 1} = 2\sqrt{\frac{I}{I_0}} \int_0^{z_e} \frac{dz}{(2\gamma'z + \gamma'^2 z^2)^{3/4}} \tag{16}$$

Where $\beta\gamma = \sqrt{\gamma^2 - 1}$ and $\gamma(z) = 1 + \gamma' z$ have been used to obtain the RHS. Once more it is useful to express the integral in terms of a dimensionless variable. Defining that variable to be $u \equiv \gamma' z$ gives

$$\int_0^{z_e} \frac{dz}{(2\gamma'z + \gamma'^2 z^2)^{3/4}} = \frac{1}{\gamma'} \int_0^{\gamma' z_e} \frac{du}{(2u + u^2)^{3/4}} \tag{17}$$

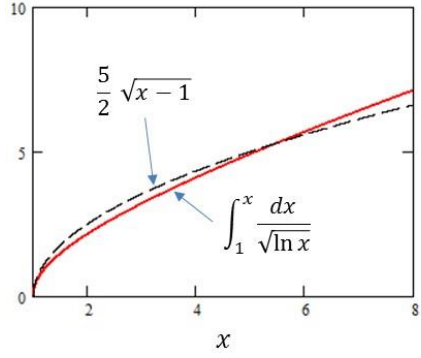

Figure 5(color): Comparison of the numerical evaluation of the dimensionless integral with the square root approximation.

Since there appears to be no known analytic solution for this integral, it is argued that its numerical solution is reasonably well approximated by the 4$^{th}$ root of $u$,

$$\int_0^u \frac{du}{(2u+u^2)^{3/4}} \cong 2(u)^{1/4} \tag{18}$$

The level of agreement between this approximation and the exact (numerical) integral is illustrated in Figure 6. Figure 6 indicates that the approximation is reasonable up to $u \sim 10$. Therefore, if $\gamma' = 40$ this approximation is good out to $z = 10/40\ m = 25\ cm$. Since the active length of the UHF gun is much less at $4\ cm$ or $u = 1.6$, therefore, using this function instead of the more complicated integral is a reasonably valid simplification. However, the approximation begins to fail for high-field guns. For example, the LCLS-I gun operates at $\gamma' = 100$ and the gun's active length is 0.12 m, therefore with $\gamma'z = 12$ which pushes the limits of these approximate functions. Although the approximation for the integral can certainly be improved, this paper will use the $2u^{1/4}$ approximation, since it captures many of the important effects occurring during the beam's acceleration from rest and is mathematically simple.

And finally, after putting it all together and solving for $r_m$, the beam envelope radius as a function of distance from the cathode is found to be

$$r_m = r_{m,0}\left(1 + \left(\frac{8}{5}\right)^2 \frac{I}{I_0} \frac{(\gamma'z_e)^{1/2}}{\gamma'^2 r_{m,0}^2}\right) \tag{19}$$

Therefore, the angle an electron at the envelope radius makes with the z-axis is

$$r'_m = \left(\frac{8}{5}\right)^2 \frac{I}{2I_0} \frac{1}{\gamma' r_{m,0} (\gamma'z_e)^{1/2}} \tag{20}$$

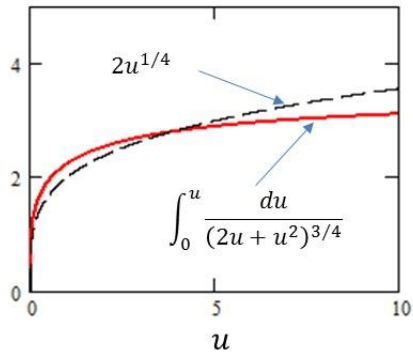

Figure 6(color): Comparison of the numerically solved integral (red-solid) with the approximate function $2(u)^{1/4}$ (black-dash).

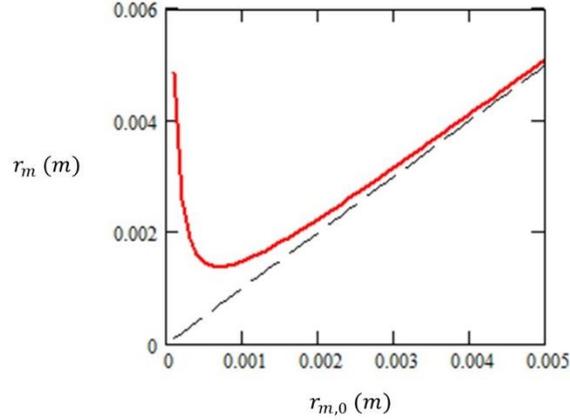

Figure 7(color): $r_m$ vs. $r_{m,0}$ as given by Eqn. (19). The red curve is computed for $\gamma' = 40, I = 4\ amperes$, and $z_e = 0.040\ m$. The dashed curve is linear with unit slope.

It is useful to plot the envelope radius at the exit of the gun vs. the beginning radius at the cathode. Such a plot of $r_m\ vs.\ r_{m,0}$ is shown in Figure 7 for $\gamma' = 40, I = 4\ amperes$ and $z_e = 0.040\ m$. Which are APEX-like parameters. The calculation shows that for large initial envelope radii, the exit envelope approaches the value of the initial radius. This is because the space-charge forces become negligibility small at large beam sizes, and since there are no other focusing or defocusing fields in this model, the beam drifts without growing larger. The difference between initial and final beam sizes falls as $1/r_{m,0}^2$ for large $r_{m,0}$. However, the envelope radius grows considerably for small values of $r_{m,0}$. This same behavior is found numerically using GPT [20] as shown later in Figure 22. This solution provides an easily quantifiable distinction between emittance-dominated and space charge-dominated beams. In the present model, the second term inside the brackets of Eqn. (19) is due to space charge forces based upon the assumptions of a radially symmetric, uniform charge density beam with constant current $I$ and radius $r_m$ in a constant longitudinal accelerating field $\gamma'$ with no transverse components.

These analytic functions for the beam's envelope as it is accelerated in the presence of space-charge forces provides scaling laws and a useful understanding of its evolution in transverse phase space, but the model also requires some physics input as well as some assumptions about the geometry to compute the emittance of the overlapping beamlets.

The normalized emittance for an uncorrelated distribution in *xx'* phase space is

$$\epsilon_x = \sigma_x \frac{\sqrt{\langle p_x^2 \rangle}}{mc} \tag{21}$$

Here it is assumed that the electrons diverge radially from the center of each beamlet and the emission from the finely distributed beamlets is all the same. Therefore, the emittance can be written as the divergence of each beamlet uniformly distributed across the full beam area times the full beam size. This same assumption is used to compute the intrinsic emittance. Thus, the beamlet values/parameters for the divergence and the full beam parameter/beam x-rms will be used below when deriving the emittance. If in addition, it is assumed that the distributions in both $r$ and $p_r$ are uniform, then the rms-values of their $x$ and $p_x$ distributions can be written as

$$\sigma_x = \langle x^2 \rangle^{1/2} = \frac{r_{m,0}}{2} \quad \text{and} \quad \langle p_x^2 \rangle = \frac{p_r^2}{4}. \tag{22}$$

Given Eqn. (20) and $\beta\gamma = \sqrt{\gamma^2 - 1}$, the radial momentum $p_r = \beta\gamma mc\, r_m'$ of an electron at the beamlet envelope is easily written as a function of $z_e$,

$$\frac{p_{r;beamlet}}{mc} = \left(\frac{8}{5}\right)^2 \frac{I_{beamlet}}{2I_0} \frac{\sqrt{2+\gamma' z_e}}{\gamma'\, r_{m,0;beamlet}} \quad (23)$$

The mesh pattern previously shown in Figure 4 has $n_s$ beamlets or current modulations across the beam diameter. Figure 8 provides more detail of the pattern showing the beamlet center-to-center spacing is assumed to be 4-times the beamlet radius. Relating this beamlet spacing with the modulation period gives the initial beamlet radius in terms of the full beam envelope radius,

$$r_{m,0;beamlet} = \frac{r_{m,0}}{2n_s} \quad (24)$$

Since the full beam current of all the beamlets is $I$, the current of a single beamlet would be $I$ divided by the number of beamlets. And the number of beamlets is just the full beam area in units of $n_s$ which is

$$N_{beamlets} = \frac{\pi}{4} n_s^2 \quad (25)$$

Therefore, the peak current of each beamlet scales as the inverse of the modulation spatial frequency squared,

$$I_{beamlet} = \frac{4}{\pi n_s^2} I \quad (26)$$

The x-plane emittance is then found from the following chain of relations,

$$\sigma_x \frac{\langle p_x^2 \rangle^{1/2}}{mc} = \sigma_{x;full} \frac{p_{r;beamlet}}{2mc} = \frac{8}{5^2} \frac{I_{beamlet}}{I_0} \frac{\sqrt{2+\gamma' z_e}}{\gamma'\, r_{m,0;beamlet}} r_{m,0} \quad (27)$$

And with the help of the relations for the beamlet envelope radius and current, the emittance is found to be the rather simple expression,

$$\epsilon_{x,sc} = \frac{1}{2\pi}\left(\frac{8}{5}\right)^2 \frac{I}{I_0} \frac{\sqrt{2+\gamma' z_e}}{\gamma' n_s} \quad (28)$$

Since the beamlets have all overlapped and the forces washed out by the smearing of the charges long before $\gamma' z_e$ is ever close to 2, the $\gamma' z_e$-term inside the radical can usually be ignored, and the emittance due to s-c of a rectangular mesh of beamlets becomes

$$\epsilon_{x,sc-mesh} = \frac{1}{\sqrt{2}\pi}\left(\frac{8}{5}\right)^2 \frac{I}{I_0} \frac{1}{\gamma' n_s} \quad (29)$$

It is worth noting that neither of these last two expressions for the emittance depend upon the beam size except through $n_s$, the number of spatial periods across the diameter. And both equations scale linearly with the current, which is also observed experimentally [15]. In addition, the mesh's space-charge emittance decreases as the inverse of the product of acceleration and the spatial frequency, $\gamma' n_s$. Thus, higher cathode fields reduce this emittance as the inverse of the field, and the low spatial frequencies are more importance than the higher spatial frequencies.

This expression for the emittance can be compared with experiments performed at the LCLS photocathode injector. In these beam studies, two very different mesh size screens were placed in the drive laser beam and imaged onto the photocathode of a high-field, 1.6-cell, s-band gun to produce beams with each mesh pattern. The experimental emittances and their analysis are given along with images of the virtual cathode in Ref. [17]. This earlier paper presents a non-relativistic version of this analysis and although it gives the correct magnitude for the emittance, unfortunately, the emittance dependences upon the beam current and size are wrong and should be replaced with the above results.

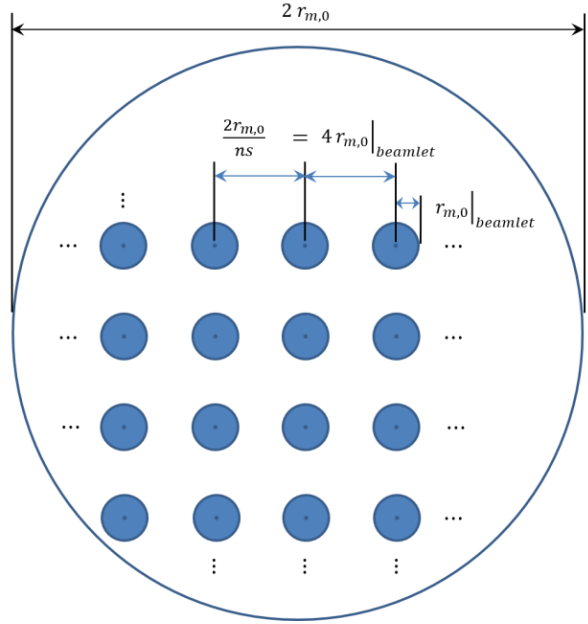

Figure 8(color): The regular mesh pattern used to compute the emittance consists of rectangular array of equally spaced beamlets within the full beam diameter of $2r_{m,0}$. The array spatial frequency is $n_s$ beamlets per full beam diameter and the beamlet centers are separated by 4-times the beamlet radius.

Figure 9 shows a plot of the theoretical emittance as a function of distance from the cathode indicates a nearly constant emittance out to a few mm's from the cathode for the mesh patterns. The figure shows, the beam envelope radius increases three or more times in this distance mixing the beamlet charge distributions in real space to give a globally (i.e., on a scale of the cathode radius) uniform charge density. This charge uniformity turns off the space-charge (s-c) force and ends s-c emittance growth of the mesh. At this point the electrons drift with constant radial velocity, and hence constant emittance. Where this transition occurs and how quickly it occurs it is determined by the beam's acceleration. For the meshes shown in the figure, the beamlet emittance due to s-c should stop growing for $z_e > 1 \, mm$ due to complete overlap. Hence the mesh emittance becomes whatever value it has at that $z_e$ where mixing is complete.

Closer to the cathode, the emittance suddenly jumps to a non-zero constant value produced instantaneously by the radial space-charge field when the beam is born. Eqn. (20) shows the envelope angle diverges at the cathode $z_e = 0$. Fortunately, it diverges slowly enough (as $1/\sqrt{z_e}$) that when multiplied by $\beta\gamma$ to normalize the emittance produces a finite and constant emittance. Eqn. (29) gives this instantaneous emittance jump due to combined radial s-c forces and longitudinal acceleration at the cathode.

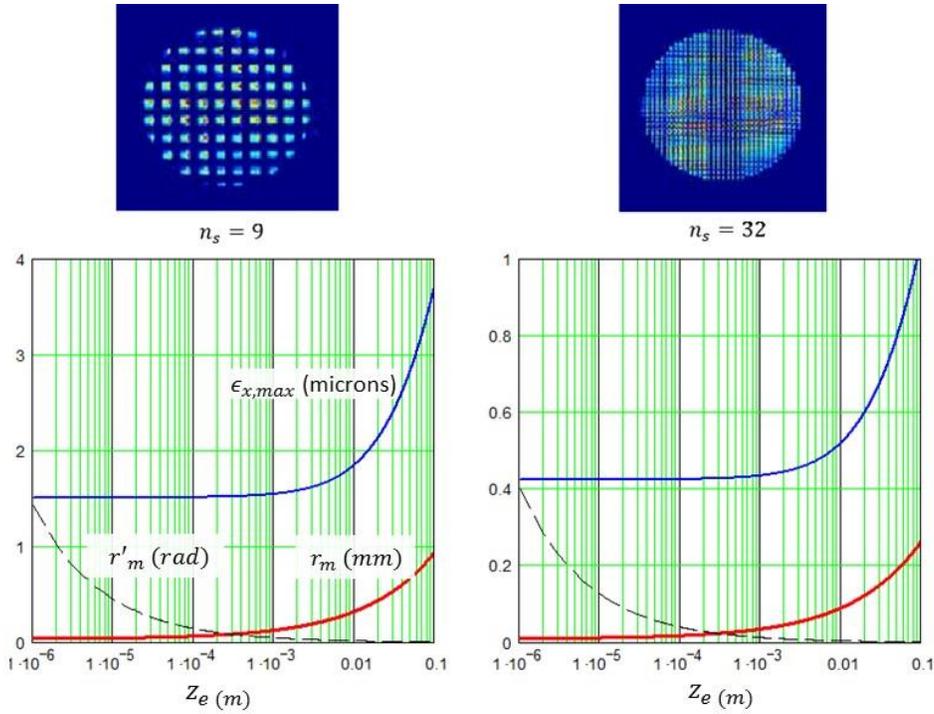

Figure 9 (color): Space-charge mesh emittance, beamlet envelope radius and envelope angle vs. distance from the cathode for two rectangular mesh spatial frequencies. The full beam parameters used in the theory are $I = 40\ amperes$, $2r_{m,0} = 1.2\ mm$, and $\gamma' = 100$, which are the same parameters for the experimental results given in Figure 10. The left and right sides of the figure are for different values for the modulation periods, $n_s$, across the full beam diameter, $2r_{m,0}$. Top: Virtual cathode images of mesh with $n_s = 9$ (left) and $n_s = 32$ (right) beamlets across the laser diameter measured for the data shown. Bottom: The beamlet envelope radius as a function of distance from the cathode is plotted with a solid-red line and the emittance as given by Eqn. (28) is shown with a solid-blue line; Right: Same as the left except with $n_s = 32$.

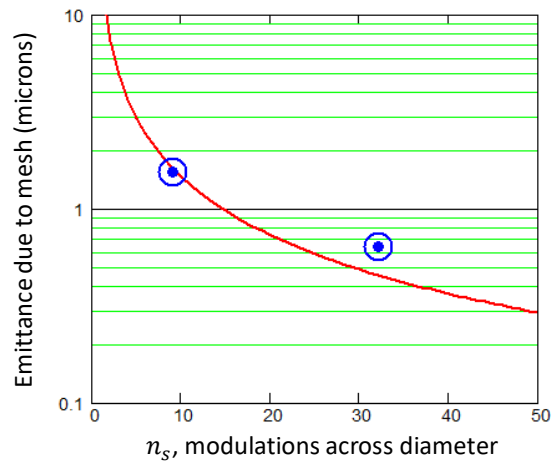

Figure 10 (color): The experimental points are from Ref. [17] and are the measured values of the emittance with the emittance due to a uniform laser beam profile subtracted off in quadrature. The nominal emittance for a uniform beam is measured to be 0.45 microns.

**RF Emittance**

The RF emittance is the projected emittance due to the time-dependent RF lens of the gun and is minimized by having the bunch on crest at the exit of the last gun cell as well as by balancing the cell-to-cell RF field amplitudes. If one assumes the length of the iris between cells is short, then the beam size is the same at both the exit and entrance of neighboring cells. In this case, for a cell-to-cell π phase shift, the defocusing field at the exit of each cell is cancelled by the entrance focus of the next cell. However, the last cell's exit field is not cancelled leaving a strong, time-dependent RF lens at the exit of the gun. This time-varying lens changes each slice's divergence along the bunch producing a projected emittance.

The total emittance can be expanded in powers of the rms bunch length, $\sigma_\phi$, and combined as the sum of the squares of the first-order and second-order RF emittances. The total RF emittance is given as,

$$\epsilon_{n,rf} = \sqrt{\epsilon_{1st}^2 + \epsilon_{2nd}^2} \tag{30}$$

The first- and second-order emittances have been computed by Kim[21] which can be summed in quadrature to give the total RF emittance,

$$\epsilon_{n,rf} = \frac{eE_{rf}}{2mc^2} \sigma_x^2 \sigma_\phi \sqrt{\cos^2\phi_e + \frac{\sigma_\phi^2}{2}\sin^2\phi_e} \tag{31}$$

Here $\sigma_x$ is the rms beam size and $\sigma_\phi$ is the rms bunch length in radians at the RF frequency. Both are evaluated at the exit of the gun where the bunch-rf phase is given by $\phi_e$. $E_{rf}$ is the peak RF field of the gun and $\phi_e$ is the electron phase relative to the RF waveform when the electron bunch reaches the exit of the gun. The total, first- and second-order projected RF emittances as functions of the exit phase are shown in Figure 11. The plots indicate that there is always a RF emittance which even at the minimum of the linear term is where the second-order term is a maximum. Eqn. (31) shows the second-order emittance grows as the square of the bunch length which in practice limits the operating bunch length to approximately ten degrees of RF phase. The second-order emittance can be eliminated by adding a third harmonic of the RF field in a two-frequency RF gun [22].

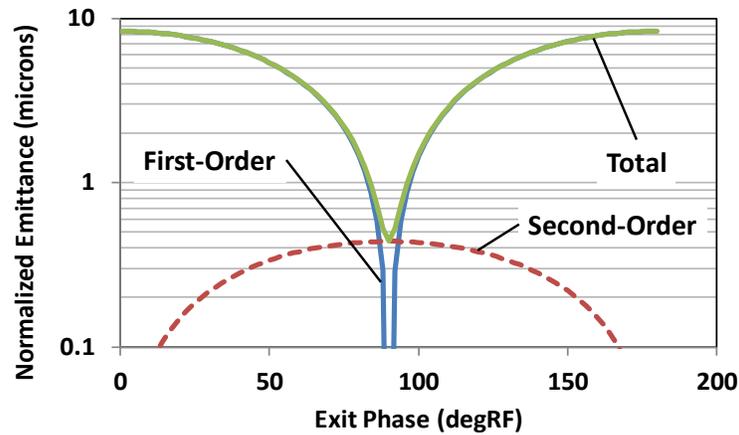

Figure 11 (color): The RF emittance as a function of the exit phase for a 100 MV/m gun with a 1 mm rms size beam and a Gaussian longitudinal distribution of 4 degrees rms at the exit iris. The total emittance (green solid) is the sum of the first-order (red solid) and the second-order (blue dash) emittances.

Table I lists the RF emittances and the measured projected emittances at 1 nC, 250 and 20 pC for the LCLS injector [23]. The RF emittances are computed using Eq. (31) for LCLS s-band (2.856 GHz) gun parameters of $E_0 = 115$ MV/m, $\phi_e = \pi/2$, $\sigma_{x,e} = 1$ mm. The rms phase length $\sigma_{\phi,e}$ given in the table was measured with a transverse deflecting RF cavity at 135 MeV. The RF emittance is computed using the experimental bunch lengths at 20 pC, 250 pC and 1 nC. The experimental projected emittance is significantly higher and shown to illustrate the RF emittance is a small contributor to other emittance sources. The magnitude of the RF emittance relative to the other emittances and the total emittance is discussed later in the paper.

Table I
The measured 1 nC, 250 pC and 20 pC electron bunch properties and the RF emittance given by Eqn. (31)

| Charge | 1 nC | 250 pC | 20 pC |
|---|---|---|---|
| Bunch length, expt (mm-rms) | 1.10 | 0.74 | 0.21 |
| Phase length, $\sigma_{\phi,e}$ expt (rad-rms) | 0.064 | 0.043 | 0.012 |
| RF Emittance at $\phi_e = \frac{\pi}{2}$, $\varepsilon_{rf}$ (microns) | 0.33 | 0.15 | 0.011 |
| Projected Emittance, expt. (microns) [22] | 1.2 | 0.70 | 0.14 |

**Chromatic Aberration of the Gun Solenoid**

Due to the strong defocusing of the RF gun it is necessary to use a comparably strong focusing lens to collimate and match the beam into the high-energy booster linac. If this focusing is done with a solenoid, then its focal strength in the rotating frame of the electrons is [24]

$$\frac{1}{f_{sol}} = K \sin KL \quad \text{with} \quad K \equiv \frac{eB(0)}{2p} = \frac{B(0)}{2(B\rho)_0} \tag{32}$$

Where $B(0)$ is the peak interior field of the solenoid, $L$ is the solenoid effective length, $(B\rho)_0$ is the magnetic rigidity, $e$ is the electron charge and $p$ is the beam momentum. The rigidity can be expressed in the following useful units as

$$(B\rho)_0 = 33.356 \, p\left(\frac{GeV}{c}\right) kG \cdot m \tag{33}$$

with $p$ being the electron momentum. It can be shown that the normalized emittance due to the chromatic aberration of a lens is [25, 26]

$$\epsilon_{n,chromatic} = \beta\gamma\sigma_{x,sol}^2 \sigma_p \left|\frac{d}{dp}\left(\frac{1}{f_{sol}}\right)\right| . \tag{34}$$

Here $\beta$ is the beam velocity divided by the speed of light, $\gamma$ is the beam's Lorentz factor, $\sigma_{x,sol}$ is the transverse rms beam size at the entrance to the solenoid and $\sigma_p$ is the rms momentum spread of the beam. Using Eqn. (32) in Eqn. (34) gives the chromatic emittance as

$$\epsilon_{n,chromatic} = \sigma_{x,sol}^2 \frac{\sigma_p}{mc} K|\sin KL + KL \cos KL| . \tag{35}$$

Figure 12 is a plot of the chromatic emittance of the solenoid as a function of the rms energy spread as given by Eqn. (35) and as simulated by GPT [20]. The beam kinetic energy is 6 MeV and the solenoid effective length is 19.35 cm with a field of 2.4 kG. The initial beam had zero emittance (zero divergence) with a 1 mm-rms transverse beam size. There are similar conditions assumed in the above derivation. There is excellent agreement between the analytic and numerical approaches. Both Eqn. (35) and the simulation assume the initial beam has zero emittance and is perfectly collimated going into the solenoid.

Ranges for typical bunch projected and slice electron energy spreads show the projected chromatic emittance is ~0.3 microns and the slice chromatic emittance is 0.02 to 0.03 microns. The LCLS projected emittance measured at 250 pC is 0.7 microns.

While the solenoid's chromatic aberration can be a significant part of the projected emittance, its contribution is much less for the slice emittance due to its small slice energy spread of less than a keV. Thus, the chromatic emittance for a slice is only ~0.02 microns/mm-rms. It is also important to note that since the beam size at the solenoid lens enters to the second power in Eqn. (35), the solenoid's chromaticity can introduce considerable emittance if the beam size at the solenoid is large. In practice the beam size at the solenoid varies widely with the cathode size and the bunch charge which strongly influences the projected emittance. These effects are discussed in later sections of the paper.

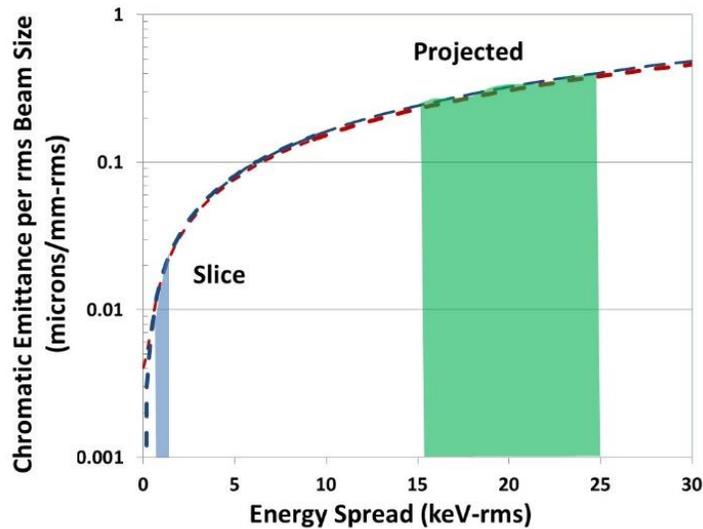

Figure 12 (color): Comparison of the chromatic trace space emittance given by Eqn. (35) (dashed blue) and the emittance computed using the GPT particle pusher code (dashed red) vs. rms energy spread. The shaded regions indicate the range of rms energy spreads within a slice (blue) and over the entire bunch (green).

**The Solenoid's Geometric Aberration**

All magnetic field solenoids exhibit a 3$^{rd}$ order angular aberration also known as the spherical aberration in classic light optics. The fields producing this aberration are dominantly located at the ends of the solenoid. This is because the aberration depends upon the second derivative of the axial field with respect to the beam direction [27]. While in theory the spherical aberration could be computed directly from the solenoid's magnetic field, in practice this is difficult and doesn't account for all the important details of the beam dynamics. Therefore, to numerically isolate the geometrical aberration from other effects, a simulation was performed with only the solenoid followed by a simple drift. Maxwell's equations were used to extrapolate the measured axial magnetic field, $B_z(z)$, to obtain the radial fields [28]. The axial field is shown below in Figure 15. Following tradition, the aberration is illustrated using an initial beam square, 2 mm × 2 mm, distribution. The simulation assumed perfect collimation (zero divergence = zero emittance), zero energy spread and an energy of 6 MeV. The transverse beam profiles given in Figure 13 show how an otherwise "perfect" solenoid produces the characteristic "pincushion" distortion [29]. A 4 mm × 4 mm (edge-to-edge) object gives 0.01micron rms emittance, while 2 mm × 2 mm square results in only 0.0025 microns.

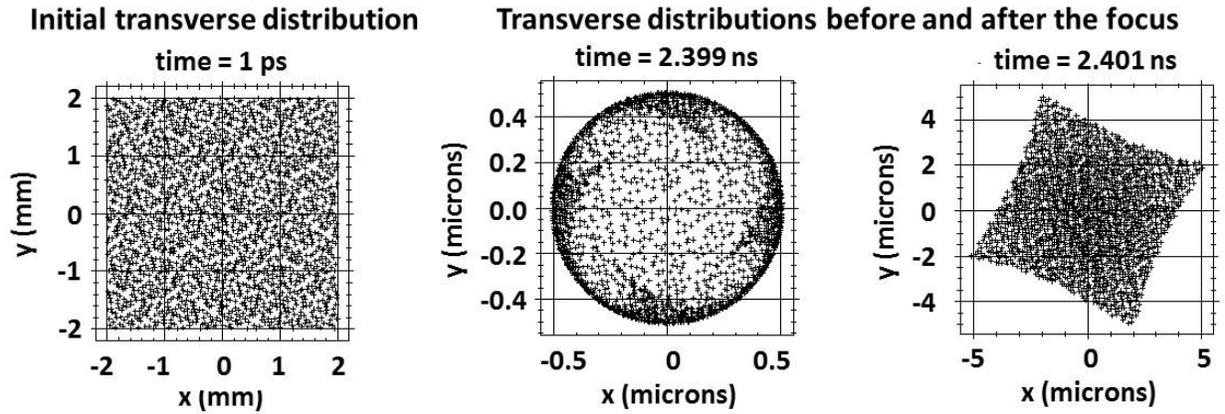

Figure 13: Ray tracing simulation of the geometric aberration of the LCLS gun solenoid. Left: the initial transverse particle distribution before the solenoid with zero emittance and energy spread. Center: The transverse beam distribution occurring slightly before the beam focus after the solenoid illustrating the third-order distortion. Right: The beam distribution immediately after the beam focus showing the third-order distortion evolving into the iconic "pincushion" shape of the rotated geometric aberration.

Figure 14 plots the simulated emittance due to the geometric aberration as a function of rms beam size at the entrance of the solenoid for an initially uniform, circular beam with an initial zero emittance. The initial beam has an energy of 6 MeV with zero energy spread. The points are the simulation and the green curve gives the 4$^{th}$ order polynomial fit.

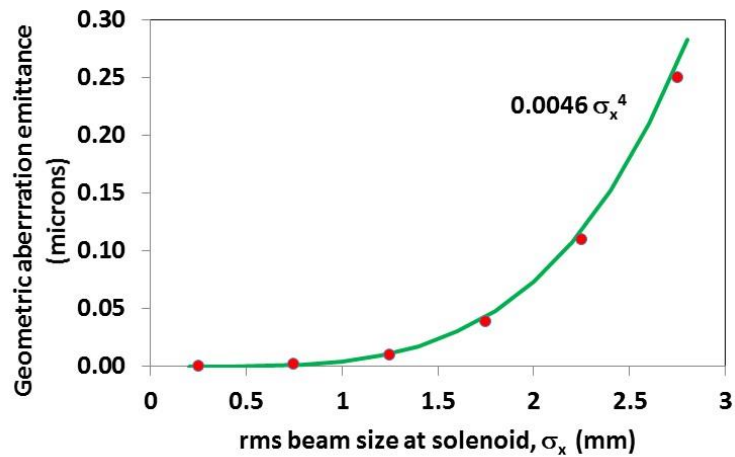

Figure 14(color): The geometric aberration for the gun solenoid: emittance vs. the x-rms beam size at the lens. The emittance computed with GPT (points red) compared with a fourth order fit (solid green). The simulation used the axial magnetic field (shown below in Figure 15) measured for the LCLS solenoid. The initial beam has zero emittance and zero energy spread.

**Anomalous or Stray Quadrupole Fields in a Solenoid Magnet**

Beam studies at the SSRL Gun Test Facility (GTF) showed the beam was astigmatic (unequal *x*- and *y*-plane focusing) which was due either to the single-side RF feed or to the magnetic field asymmetries of the gun solenoid. To understand and distinguish between these effects, the solenoid's multipole magnetic field was measured using a rotating coil. The magnetic measurements showed small quadrupole fields at the ends of the solenoid with equivalent focal lengths at 6 MeV of 20 to 30 meters for the GTF solenoid. However even though these fields were weak, it was decided to install normal and skew quadrupole correctors inside the bore of the solenoid to correct them. As described below, beam measurements show these correctors have a relatively strong influence on the emittance. Technical details of why and how the correctors were incorporated into the gun solenoid are given in Ref. [30] and their use during operation is described in Ref. [23]. This section discusses the dynamics of a beam in combined axial and quadrupole magnetic fields. The interested reader is directed to Ref. [26] for further details.

Figure 15 shows the axial magnetic field and the quadrupole magnetic field and its angular orientation or phase angle along the beam axis of the LCLS solenoid. The quadrupole field was measured using a rotating coil which was 2.5 cm long with a 2.8 cm radius. This is the radius for which the quadrupole field is given in the figure. The quadrupole phase angle is the angular rotation of the poles relative to an aligned quadrupole, and is the angle of the quadrupole north pole relative to the *y*-axis (left when travelling in the beam direction) for beam-centric, right-handed coordinate system. In this coordinate system, a normally aligned quadrupole has a phase angle of 45 degrees. The difference in phase angle between the entrance ($z = -9.6$ cm) quadrupole field and the exit ($z = +9.6$ cm) field angle is close to 90 degrees. Thus, these anomalous end fields have opposing polarities which reverse sign when the solenoid's polarity is reversed. These LCLS solenoid fields are qualitatively like those measured previously for the GTF solenoid, although the overall magnitude of the fields is lower. The LCLS solenoid has an equivalent focal length of approximately 50 meters due to these small quadrupole fields while the GTF solenoid's anomalous quadrupole fields corresponded to 20 to 30 meters.

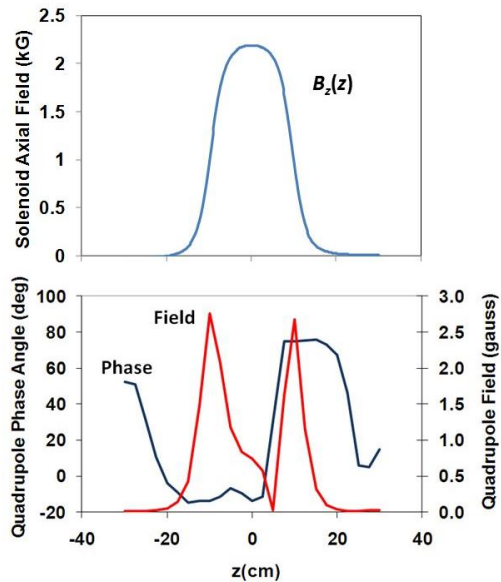

Figure 15(color): Magnetic measurements of the LCLS gun solenoid. Top: Hall probe measurements of the solenoid axial field. The transverse location of the measurement axis (the *z*-axis) was determined by minimizing the dipole field. Bottom: Rotating coil measurements of the quadrupole field. The rotating coil dimensions were 2.5 cm long with a 2.8 cm radius. The measured quadrupole field is thus averaged over these dimensions.

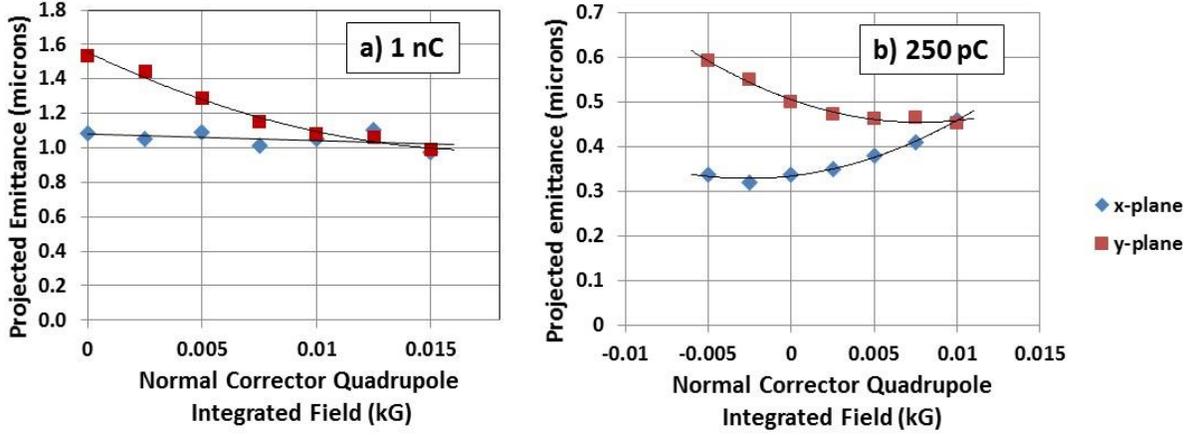

Figure 16: a) Measured *x*-plane (blue) and *y*-plane (green) emittances vs. the normal corrector quadrupole strength for a 1 nC bunch charge. b) Similar measurement for 250 pC. The lines are shown to guide the eye.

As described earlier, the correction of these anomalous quadrupole fields was done by installing normal and skew quadrupoles inside the bore of the solenoid. The effect these correction quadrupoles have upon the emittance is quite profound, as can be seen in Figure 16 where the measured emittance for 1 nC and 250 pC are plotted vs. the normal corrector quadrupole strength.

**The Emittance due to the Anomalous Quadrupole Fields**

The beam emittance due to these anomalous quadrupole fields can be computed both in simulation and analytically. The analysis begins by assuming a simple thin quadrupole lens followed by a solenoid with the 4×4 *x-y* beam coordinate transformation given by

$$R_{sol}R_{quad} = \begin{pmatrix} \cos^2 KL & \dfrac{\sin KL \cos KL}{K} & \sin KL \cos KL & \dfrac{\sin^2 KL}{K} \\ -K \sin KL \cos KL & \cos^2 KL & -K \sin^2 KL & \sin KL \cos KL \\ -\sin KL \cos KL & -\dfrac{\sin^2 KL}{K} & \cos^2 KL & \dfrac{\sin KL \cos KL}{K} \\ K \sin^2 KL & -\sin KL \cos KL & -K \sin KL \cos KL & \cos^2 KL \end{pmatrix} \begin{pmatrix} 1 & 0 & 0 & 0 \\ -\dfrac{1}{f_q} & 1 & 0 & 0 \\ 0 & 0 & 1 & 0 \\ 0 & 0 & \dfrac{1}{f_q} & 1 \end{pmatrix} \quad (36)$$

As in the derivation for the chromatic emittance: $L$ is the effective length of the solenoid, $K \equiv \frac{eB_z(0)}{2p}$, $B_z(0)$ is the interior peak axial magnetic field of the solenoid, and $f_q$ is the focal length of the anomalous quadrupole field located at the entrance to the solenoid. The beam is rotated through the angle $KL$ by the solenoid.

The 4×4 covariance matrix of the beam after the combined quadrupole and solenoid is then

$$\sigma(1) = R_{sol}R_{quad}\,\sigma(0)\,\left(R_{sol}R_{quad}\right)^T \quad (37)$$

with the *x*-plane emittance after the quadrupole and solenoid being given by the determinate of the 2×2 sub-matrix,

$$\epsilon_{n,qs} = \beta\gamma\sqrt{\det \sigma_x(1)} = \beta\gamma\sqrt{\det\begin{pmatrix}\sigma_{11}(1) & \sigma_{12}(1)\\ \sigma_{12}(1) & \sigma_{11}(1)\end{pmatrix}} \tag{38}$$

And finally, the normalized emittance due to an anomalous quadrupole field near the solenoid entrance is found to be

$$\epsilon_{n,qs} = \beta\gamma\sigma_{x,sol}\sigma_{y,sol}\left|\frac{\sin 2KL}{f_q}\right| \tag{39}$$

The x and y transverse rms beam sizes are the entrance to the solenoid are $\sigma_{x,sol}$ and $\sigma_{y,sol}$.

Figure 17 compares this simple formula with a particle tracking simulation as done for the geometric aberration. In this case the simulation was done for a solenoid followed by a drift with a weak quadrupole field overlapping the solenoid field. The initial beam had zero emittance, zero energy spread and was circular and uniform. No space charge forces are included in the simulation. Figure 17 shows the normalized emittances given by Eqn. (39) and the simulation plotted as a function of the rms beam size at the solenoid entrance. The anomalous quadrupole focal length is 50 meters at 6 MeV which is approximately the same as computed from the magnetic measurements for the LCLS solenoid. Both the analytic theory and the simulation assume a short quadrupole field only at the solenoid's entrance. The simulation is slightly larger since includes both this quadrupole effect and the geometric aberration described above. The good agreement verifies the model's basic assumptions and illustrates how even a very weak quadrupole field can strongly affect the emittance when combined with the rotation in a solenoid field.

Eqn. (39) is for a quadrupole plus solenoid where the anomalous quadrupole field isn't rotated with respect to a normally oriented quadrupole. When the quadrupole is rotated about the beam axis by angle, $\alpha$, then the effective rotation becomes the sum of the quadrupole rotation plus the beam rotation in the solenoid and the emittance becomes

$$\epsilon_{n,qs}(\alpha) = \beta\gamma\sigma_{x,sol}\sigma_{y,sol}\left|\frac{\sin 2(KL+\alpha)}{f_q}\right| \tag{40}$$

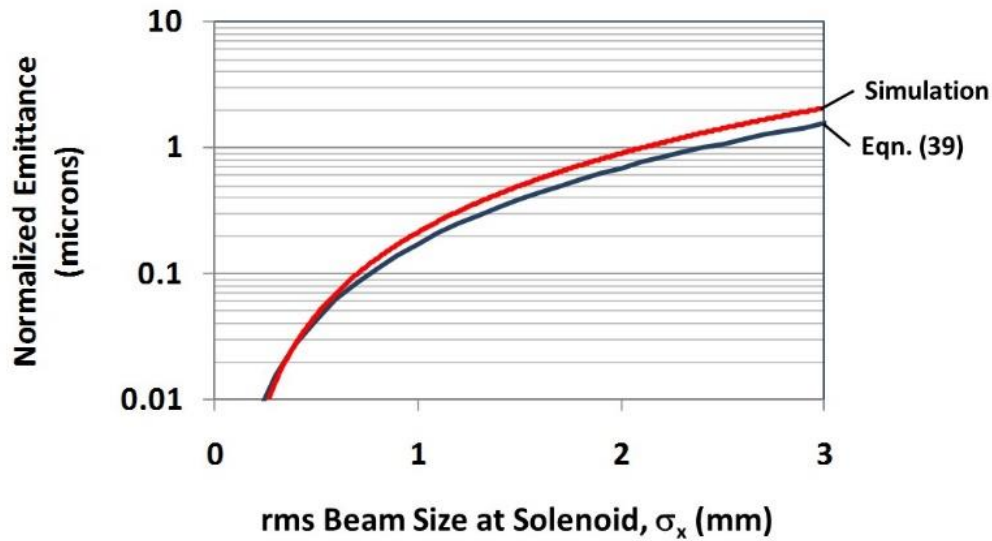

Figure 17(color): Comparison of the emittance due to the quadrupole-solenoid coupling given by Eqn. (23) with a particle tracking simulation for the case of the LCLS solenoid. For a beam energy of 6 MeV the quadrupole focal length was 50 meters and the solenoid had an integrated field of 0.46 kG-m.

Figure 18 compares Eqn. (40) with a simulation for a 50-meter focal length quadrupole followed by a strong solenoid (focal length of ~15 cm). The emittance is plotted as a function of the quadrupole angle of rotation. In both the analytic theory and the simulation, the emittance becomes zero when $KL + \alpha = n\pi$. The slight shift in angle between the theory and simulation occurs because the simulated solenoid has fringe fields which are ignored in the theory.

To model the full effect of skewed quadrupole field errors at both ends of the solenoid it is necessary to express the emittance for a quadrupole pair with rotation angles of $\alpha_1$ and $\alpha_2$, and focal lengths of $f_1$ and $f_2$, respectively. Following the same procedure used to derive Eqn. (39) one finds,

$$\epsilon_{n,qs}(\alpha_1, \alpha_2) = \beta\gamma\sigma_{x,sol}\sigma_{y,sol} \left| \frac{\sin 2\alpha_1}{f_1} + \frac{\sin 2\alpha_2}{f_2} \right| \quad (41)$$

Here it is assumed the beam size at the quadrupoles is the same as at the solenoid. The final emittance is due to three effects: the skew angle and focal length of the entrance quadrupole, $(\alpha_1, f_1)$, the rotation in the solenoid, $KL$, and the skew angle and focal length of the exit quadrupole $(\alpha_2, f_2)$. Combining the entrance quadrupole skew angle with the solenoid rotation, one obtains the emittance for a solenoid with anomalous quadrupole end fields,

$$\epsilon_{n,qs}(\alpha_1, KL, \alpha_2) = \beta\gamma\sigma_{x,sol}\sigma_{y,sol} \left| \frac{\sin 2(KL+\alpha_1)}{f_1} + \frac{\sin 2\alpha_2}{f_2} \right| \quad (42)$$

It is relevant to point out some of the features of Eqn. (42). First consider the situation when both quadrupoles are perfectly aligned without any skew, i.e. $\alpha_1 = \alpha_2 = 0$. Then while there's no emittance contribution from the exit quadrupole, the entrance quadrupole still appears skewed by the beam's rotation in the solenoid and the emittance increases unless there is no entrance quadrupole field. For this case the emittance does not depend upon the polarity of the solenoid field. However, this is not true when either $\alpha_1$ or $\alpha_2 \neq 0$. Eqn. (42) also shows that if the entrance quadrupole field is skewed, the emittance will depend upon the polarity of the solenoid field. Further details of this effect are discussed in the next section. Lastly, the formula indicates that adding skewed and normal quadrupole correctors near the solenoid can cancel this effect and completely recover the initial emittance.

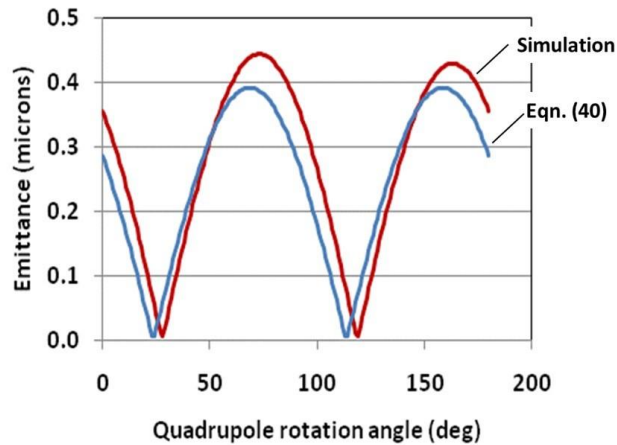

Figure 18(color): The emittance for a quadrupole-solenoid system plotted as a function of the quadrupole rotation angle. The theory emittance (solid blue) is computed using Eqn. (40) and the simulation (solid red) is done with the GPT code. The beam size at the solenoid is 1 mm-rms for both the x- and y-planes.

**Correcting the Solenoid's Anomalous Quadrupole Field Emittance**

As just described the emittance growth due to the solenoid's anomalous quadrupole fields can be compensated with the addition of skew and normal corrector quadrupoles. In the LCLS solenoid these correctors consist of eight long wires inside the solenoid field, four in a normal quadrupole configuration and four arranged with a skewed quadrupole angle of 45 degrees. Thus, since corrector quadrupoles overlap the solenoid field, one would expect their skew angles should be added to *KL* as done in the first term of Eqn. (42). The emittance due to the composite system of a rotated quadrupole in front of the solenoid, the two corrector quadrupoles inside the solenoid and the exit rotated quadrupole can be computed as the following sum,

$$\epsilon_{n,normal+skew}(\alpha_1, KL, \alpha_2) = \beta\gamma\sigma_{x,sol}\sigma_{y,sol} \left| \frac{\sin 2(KL+\alpha_1)}{f_1} + \frac{\sin 2KL}{f_{normal}} + \frac{\sin 2\left(KL+\frac{\pi}{4}\right)}{f_{skew}} + \frac{\sin 2\alpha_2}{f_2} \right| \quad (43)$$

The first and fourth terms inside the absolute-value brackets are due to the entrance and exit anomalous quadrupole fields with focal lengths $f_1$ and $f_2$ and skew angles of $\alpha_1$ and $\alpha_2$, respectively. The second and third terms are approximations for the normal and skew corrector quadrupoles with focal lengths $f_{normal}$ and $f_{skew}$, respectively. The normal and skew corrector quadrupoles are located before the solenoid and are rotated 0 and π/4, respectively, about the beam direction or z-axis. As expected, for no solenoid field $K = 0$, and there is no emittance due to the normal and skew quadrupoles. This expression illustrates how the solenoid's rotation of the beam amplifies the effect even weak quadrupole fields have upon the 2D emittance. Since the 4D phase space change is correlated and the 4D emittance remains zero, however, the projected emittances of the 2D subspaces xx' and yy' can become quite large. But in the end, the 4D emittance also increases as the correlation becomes lost in subsequent beam transport and optics.

Figure 19 illustrates the emittance due to these effects as a function of the normal and skew corrector quadrupole focal lengths using Eqn. (43). The entrance and exit anomalous quadrupole focal lengths are 50 meters and their rotation angles as indicated by Figure 15 are -15 and 75 degrees, respectively. The *x* and *y* rms beam sizes at the solenoid entrance are 1 mm. The red curves are for the normal corrector quadrupole only with the skew corrector quadrupole off, while the blue curves are given for the skew quadrupole only with the normal quadrupole off. The zero of emittance is shifted for the two correctors since the overall rotation necessary to correct the error fields is neither normal nor skewed, but something in between. Both solid curves asymptotically converge to the uncorrected emittance as the correctors are turned off (infinite focal length).

The figure also shows the effect of reversing the polarity of the solenoid with corresponding emittances plotted as dashed lines. In this case the uncorrected emittance clearly approaches a much smaller emittance. As mentioned earlier, the skewed anomalous quadrupole fields make the resulting emittance growth and focusing of the solenoid dependent upon its polarity and provide an experimental signature that the fields are skewed. Therefore, if the anomalous fields are skewed, one polarity of the solenoid results in a lower emittance than the other. This can be seen in Figure 19 in the limit of very weak (infinite focal length) correctors.

And finally, it's important noting that the skewed-normal quadrupoles can be sued to correct for field asymmetries of RF couplers. The DC quadrupole fields need to uncouple the RF's anomalous quadrupole field only at the time the beam reaches the coupler field. Therefore, the coupler field asymmetry can be nearly perfectly cancelled with a simple, weak skewed quadrupole field which remove the x-y correlation produced by the asymmetric coupler fields [31], [32].

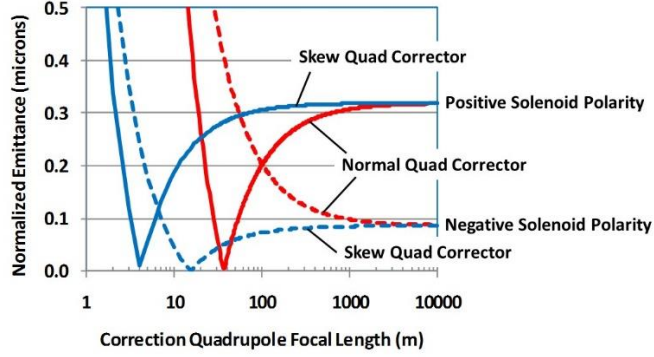

Figure 19(color): The emittance as a function of the normal (red curves) and skew (blue curves) quadrupole corrector focal lengths for positive (solid curves) and negative (dashed curves) polarities of the solenoid. The solenoid *KL* is 2.3 and *L* is 0.1935 m. Anomalous quadrupole fields with 50 meter focal lengths at the solenoid ends with quadrupole rotation or phase angles of -15 and 75 degrees, respectively, see Figure 15.

**Summing the Effects of the Solenoid**

In this section the solenoid's chromatic, geometric, and anomalous quadrupole emittances are compared as a function of the beam size at the solenoid. A general conclusion is that the emittance due to the solenoid's aberrations can be minimized with a small beam size in the solenoid. For the solenoid's geometric aberration, the emittance is proportional to the $4^{th}$ power of the beam size (see Figure 14) with the assumption that the beam is circular $\sigma_x = \sigma_y$ and equal to $\sigma_{x,sol}$ which is the transverse rms beam size at the solenoid,

$$\epsilon_{n,geometric} = 0.0046 \left(\frac{microns}{mm^4}\right) \sigma_{x,sol}^4 (mm) \qquad (44)$$

The chromatic emittance was given above as

$$\epsilon_{n,chromatic} = \beta\gamma\sigma_{x,sol}^2 K |\sin KL + KL \cos KL| \frac{\sigma_p}{p} \qquad (45)$$

And the expression for a non-skewed anomalous entrance quadrupole + solenoid emittance is assumed for a round beam,

$$\epsilon_{n,qs} = \beta\gamma\sigma_{x,sol}^2 \left|\frac{\sin 2KL}{f_q}\right| \qquad (46)$$

These three expressions are compared in Figure 20 for the 19.35 cm long LCLS solenoid using the $B_z(z)$ field profile given by magnetic measurements. The solenoid interior field is 2.6 kG. The beam total energy is 6 MeV and curves are given for rms-energy spreads of 1 and 20 keV, corresponding to typical slice and projected rms energy spreads. The anomalous quadrupole emittance is computed with a 50-meter focal length.

Figure 20 indicates solenoid's largest contribution to the emittance comes from the quadrupole-solenoid aberration. It is important to comment that 50 meters was chosen for the anomalous quadrupole focal length is to illustrate the effect. In general, the anomalous focal length is rarely measured and can vary greatly depending upon the details of the solenoid design. Fortunately, this emittance can be corrected and made essentially zero with normal and skewed correction quadrupoles. The next contributor is the chromatic aberration. The figure shows the effect for a 1 keV rms energy spread which is estimated to be the relevant slice energy spread. The contribution is much larger for the projected energy spread of ~20 keV as measured for the LCLS beam at 6 MeV. Both the chromatic and geometric aberrations, since they depend upon the beam size to the $2^{nd}$-order and $4^{th}$-order respectively, are controlled by reducing the beam size at the solenoid.

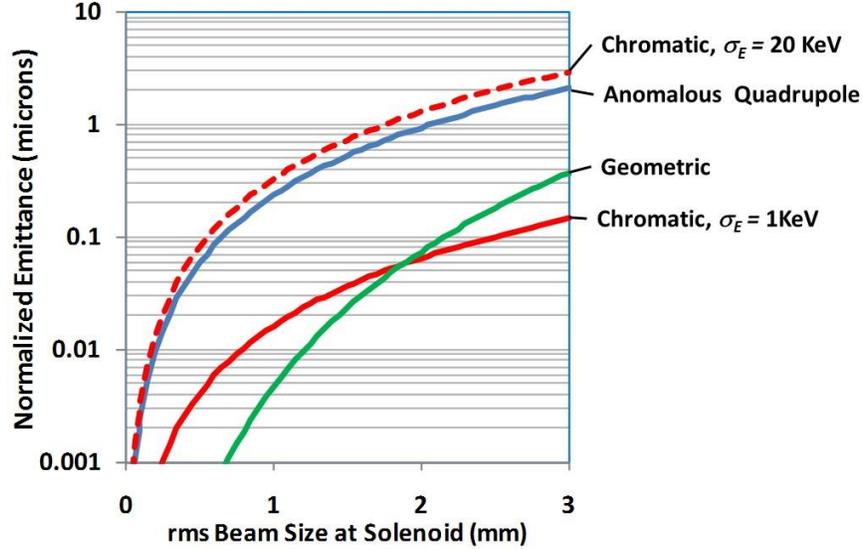

Figure 20(color): The chromatic, geometric, and anomalous quadrupole emittances as a function of the beam size at the entrance to the solenoid. The chromatic emittance is plotted for slice (1 keV) and projected (20 keV) energy spreads. The anomalous quadrupole aberration was computed using 50 m for the quadrupole focal length and the solenoid's *KL* was 1.16 with an effective length of 0.1935 m.

The anomalous quadrupole emittance can be the largest of the emittances and is due to the correlation between the x- and y-components in 2D trace space caused by the anomalous quadrupole becoming skewed by the beam's rotation in the solenoid. Thus, the trace space x- and y-emittances can be large while the 4D-emittance remains zero. This means the transverse phase space can be rotated with corrector quadrupoles to undo the correlation and correct for the anomalous quadrupole trace space emittance.

In light optics, the geometric emittance is known as the spherical aberration [29] and is present in all lenses with rotation symmetry. For electrons, the aberration is a third-order dependence of the divergence upon beam size, making the emittance of spherical aberrations scale as the 4$^{th}$ power of the beam size. In the paraxial ray equation, the there is a third order radial term whose coefficient is proportional to the sum of the field strength to the 4$^{th}$ power and the curvature of the solenoid field [18],

$$r'' \propto K^2 \left(\frac{B''}{2B} - K^2\right) r^3 + \cdots \qquad (47)$$

Therefore, the strength of the aberration depends upon the shape and strength of the magnetic field. The form of the coefficient of the *r³* term suggests that the spherical aberration of a solenoid can be minimized with proper shaping of the field at the operational field strength.

The anomalous quadrupole emittance can also be canceled with a solenoid powered such that each half has opposite axial fields. However, the chromatic emittance is unaffected by this configuration, since it is an even function of K. And as shown by Eqn. (47) the geometric aberration is also unchanged by the change in polarity and in fact is increased by the additional exit and entrance fringe fields between the two halves. However, this scheme does eliminate beam steering due to mechanical misalignment and the additional emittance due to the geometric aberration is typically small.

Eqn. (45) gives the chromatic emittance for an rms momentum spread, $\sigma_p$, due to either a *p* uncorrelated or *p-z* correlated momentum distribution. For a slice, the momentum spread is small and uncorrelated. The projected momentum spread is typically much larger and is usually due to a correlation between the momentum and the longitudinal position along the bunch, although other correlations are possible. This correlation can be one of two basic shapes in phase space. The first occurs when the bunch launch phase is set to produce a linear chip on the bunch. The case of when the bunch is being compressed in the gun is shown in the figure. The second phase space shape occurs when the launch phase is set to place the beam on the crest of the RF which gives the center electrons the highest energy with both the head and tail electrons both at lower energies.

**Beam Size at the Solenoid vs. Cathode Radius**

Estimating the solenoid emittance requires knowing the beam size at the solenoid and in most cases the beam size was assumed to be 1 mm-rms. However, in many cases the beam size at the solenoid lens is considerably larger. Figure 21 shows a simulation of the beam size from the cathode to the entrance of the LCLS booster linac, over a distance of approximately 90 cm. The legend to the right gives the bunch charge and the cathode beam size for each curve. The behavior common to all cases is the beam expands from the cathode with a small bump in the size at the gun irises near 8 cm and then is focused by the solenoid which is 19.35 cm long, centered near 30 cm from the cathode. The figure clearly shows the solenoid beam size is dependent upon both charge (1, 10, 100 and 200 pC) and the initial beam size at the cathode (0.01, 0.1, 0.3 and 0.6 mm-rms). The laser spot size on the cathode strongly affects the beam envelope. Indeed, its size at the solenoid for 1 pC can actually be larger than that of a 10 pC bunch if the laser spot is too small. The beam size at the solenoid for 200 pC is 1.8 mm-rms, per Figure 14, corresponding to an emittance of 0.3 microns from the effects discussed above. Therefore, these phenomena can be large contributors to the observed beam emittance, making it important to reduce the beam size at the solenoid in designing future guns.

Comparing the effects of intrinsic, RF, space charge, solenoid chromatic and geometric aberrations requires a consistent set of beam sizes at the cathode, gun exit and solenoid entrance. The beam sizes at the gun and solenoid are determined by the cathode radius and the space charge forces. Knowing these beam sizes as a function of the cathode radius gives a consistent analysis of the total emittance and its parts.

The beam sizes are simulated using the GPT code to model an s-band gun without and with space charge. Space charge will affect the beam emittance even with perfect laser shaping to limit the space charge emittance. The remaining linear space charge forces defocus the beam, making the beam larger at the gun exit and at the solenoid. This larger beam then increases the other non-charge dependent emittances due to RF, chromatic, geometric, and anomalous-quadrupole aberrations as described earlier. The simulation is used to obtain the beam sizes at the gun exit and the solenoid entrance resulting from the linear space charge force. These sizes are then used in the above derived formulas to obtain the emittance from the various emittance sources. This approach allows one to dissect the final emittance into its components.

The beam size was simulated for a 1.6 cell s-band gun with 115 MV/m peak field on the cathode. The laser launch phase with respect to the RF was 30 degS and the bunch charge was 250 pC. The laser pulse shape was a longitudinal Gaussian with a fwhm of 6 ps and the transverse x-distribution was circular and uniform. The simulated beam sizes at the gun exit iris and at the entrance to the solenoid were evaluated at 8 and 15 cm from the cathode, respectively. The full 3D space charge routine was used for the space charge calculations [33]. The rms beam size at these two locations was computed as a function of the cathode radius, i.e. the laser radius on the cathode. The space charge limited emission for these conditions and 250 pC occurs approximately at a cathode radius of 0.3 mm.

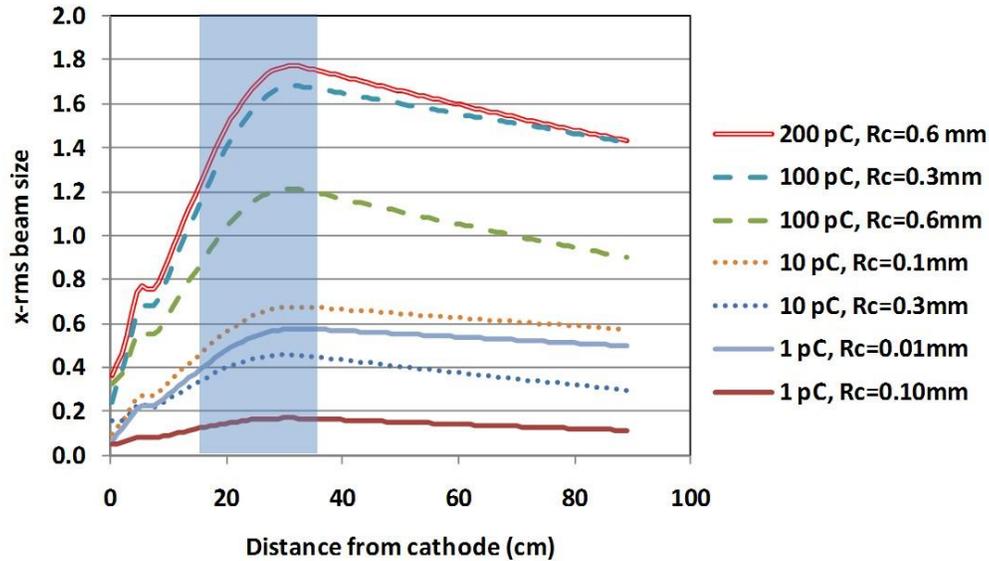

Figure 21(color): The rms beam size in millimeters as a function of distance from the cathode for a variety of beam sizes at the cathode and bunch charge. The initial transverse beam distribution is uniform with radius $R_c$. The curves are for the LCLS gun and solenoid system [23] [30] and are computed using GPT [20] [33]. The shaded region indicates the location of the solenoid axial field.

The dependence of beam size with the cathode size given in Figure 22 deserves some discussion. The curves show a minimum near a cathode radius of 0.8 mm with the size increasing more rapidly for smaller cathode radii than for larger radii. The rise in beam size below 0.8 mm is due to the linear space charge force which increases as the cathode size is reduced. A strong effect of space charge defocusing is seen for cathode radii less than ~0.8 mm. The simulation indicates that space charge defocusing adds linearly with the beam size due to the gun's RF optics. Above 0.8 mm space charge defocusing diminishes with increasing cathode radius and the beam sizes are dominated by the gun optics. A similar dependence upon size is observed in the analytic theory described earlier in this paper.

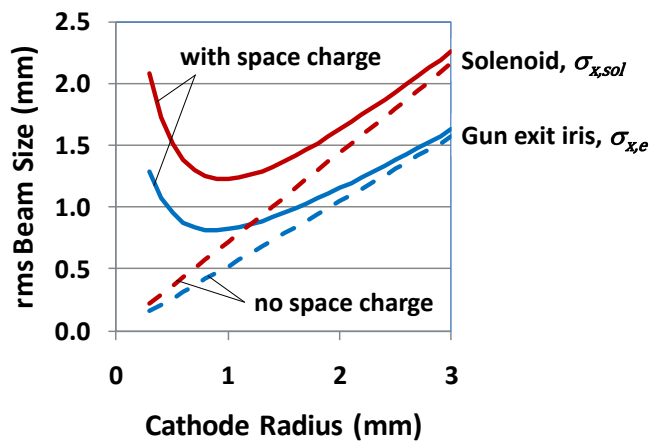

Figure 22(color): The rms beam size at the gun exit (blue) and at the solenoid entrance (red) as a function of the cathode radius. The calculation has been done without (dash) and with (solid) space charge using the 3D space charge routine in GPT [33].

**Summarizing These Effects**

In this section the emittances described in the previous sections are compared and summed for the archetypical s-band RF gun and solenoid. The emittances are re-expressed in terms of the cathode radius using the simulation results shown in Figure 22 and are summed in quadrature to obtain the total emittance. Since the emittance due to the anomalous quadrupole field is not included in this analysis since it is due to a correlation and is easily recovered.

The total emittance is defined as the square root of the quadratic sum of the five emittances described earlier,

$$\epsilon_{n,final} = \sqrt{\epsilon_{intrinsic}^2 + \epsilon_{n,sc}^2 + \epsilon_{n,rf}^2 + \epsilon_{n,chromatic}^2 + \epsilon_{n,geometric}^2} \tag{48}$$

By writing each of the five emittances in terms of the cathode radius, one can easily compare the contribution each makes to the total emittance. The beam sizes plotted in Figure 22 relate the gun exit and solenoid entrance beams sizes to the cathode radius and are used in the expressions for the RF, chromatic and geometric emittances. The calculations are done for a bunch charge of 250 pC and nominal RF and solenoid parameters for an s-band gun.

For this analysis, the experimental intrinsic emittance is used,

$$\epsilon_{n,intrinsic} = \frac{R_{cathode}}{2} \times 0.9 \frac{microns}{mm-rms} \tag{49}$$

As discussed earlier, this is approximately twice the theoretical intrinsic emittance.

The space-charge emittance has been given as

$$\epsilon_{x,s-c} = \frac{1}{\sqrt{2\pi}} \left(\frac{8}{5}\right)^2 \frac{\Delta I}{I_0} \frac{1}{\gamma' n_s} \tag{50}$$

For the comparison assume $n_s$ equals 10 periods of modulation across the bunch diameter and a current modulation depth of ten percent of the peak current. At 250 pC the peak current is 40 amperes, thus the current modulation depth, $\Delta I$, is 4 amperes peak-to-peak.

The beam bunch is assumed to exit the gun on crest to give the minimum RF emittance,

$$\epsilon_{rf}(R_{cathode}) = \frac{eE_{cathode}}{2\sqrt{2}mc^2} \sigma_{x,exit}^2(R_{cathode})\sigma_\phi^2 \tag{51}$$

Here as in the measurements $E_0$ is 115 MV/m and the bunch length is 0.74 mm-rms ($\sigma_\phi = 0.043$ radians) as given in Table II for 250 pC. The rms beam size at the gun exit as a function of the cathode beam size is given in Figure 22.

The chromatic emittance is given by

$$\epsilon_{chromatic}(R_{cathode}) = \sigma_{x,sol}(R_{cathode})^2 \frac{\sigma_p}{mc} K_{sol}|\sin K_{sol}L_{sol} + K_{sol}L_{sol} \cos K_{sol}L_{sol}| \tag{52}$$

The chromatic emittance shown in Figure 18 assumes 20 keV/c for the rms momentum spread, $\sigma_p$. This is the projected energy spread observed at 6 MeV and 250 pC for the LCLS gun. Using the energy spread of a bunch slice Eqn. (36) gives the slice chromatic emittance. The integrated solenoid field used in the measurements was 0.464 KG-m and the effective length is 0.1935 meters. These are the parameters of the LCLS solenoid [23]. This field corresponds to a $K$ of 5.99 m$^{-1}$ and $KL$ is then 1.16. And once more, the graph in Figure 22 provides the solenoid entrance beam size with space charge as a function of the cathode radius.

Figure 9 showed the following 4$^{th}$ power fit to the simulated geometric emittance as a function of beam size at the solenoid,

$$\epsilon_{geometric}(R_{cathode}) = 0.0046\sigma_{x,sol}^4(R_{cathode}) \tag{53}$$

Where $\sigma_{x,sol}(R_{cathode})$ is in units of mm.

These five projected emittances and their combined total emittance are plotted in Figure 23 as a function of the cathode radius. The space charge beam sizes at the gun exit and the solenoid entrance given in Figure 22 are used to compute the RF, chromatic and geometric emittances. The minimum in the beam size at the gun exit and solenoid near 0.9 mm cathode radius causes a local minimum in the geometric, chromatic and RF emittances. There is no minimum for the intrinsic and space charge emittances since they occur close to the cathode, far from the gun exit and solenoid. They are linear functions of the cathode radius and are independent of the gun's optics. The total emittance has a minimum shifted to a smaller cathode radius than that shown in Figure 23 for the gun and solenoid beam sizes, 0.7 mm vs. 1 mm, respectively. The measured projected emittance at 250 pC and a cathode radius of 0.6 mm is 0.7 microns [23].

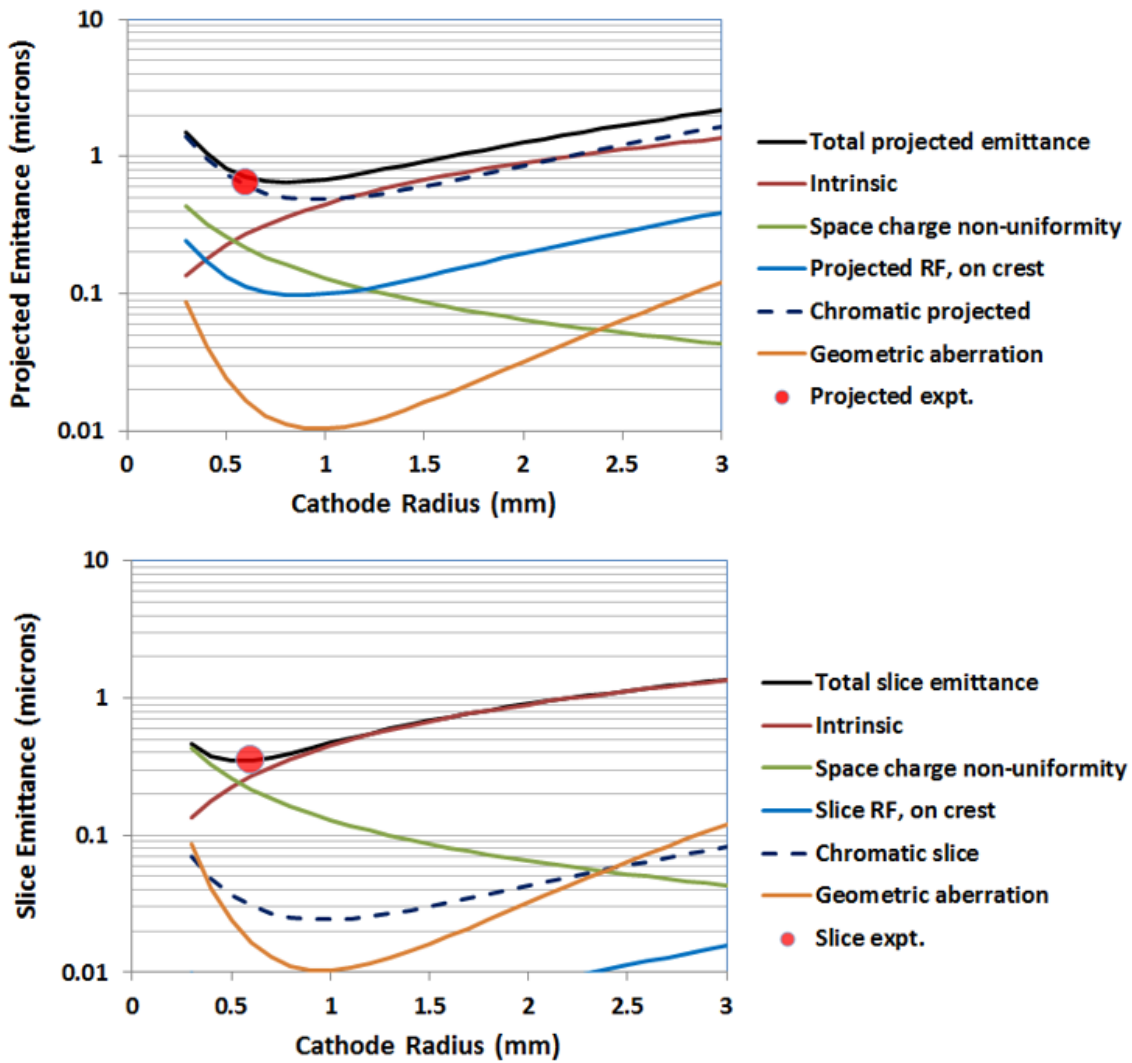

Figure 23(color): The components of the projected emittance at 250 pC bunch charge for a 1.6 cell S-band gun and emittance compensation solenoid. The red point shows the experimental projected emittance of 0.65 microns for a 0.6 mm radius cathode as measured at the LCLS injector.

**Summary and Conclusions**

This paper has attempted to identify and quantify the major sources of emittance in a photocathode RF gun and solenoid system. The principal emittances identified were the intrinsic, space charge, RF and solenoid aberration and anomalous quadrupole field emittances. The analysis used a combination of analytic and numerical techniques to derive expressions for these emittances. Simple equations for these emittances were given in terms of the cathode size, the beam size, energy spread, bunch length and charge. A comparison of these effects was done for a typical s-band gun showing that the chromatic, intrinsic and space charge emittances are the main contributors to the total projected emittance. The RF emittance is approximately five-times smaller than the intrinsic emittance. The 4$^{th}$ order geometric emittance is a decade smaller than the RF emittance. There is general agreement between this model and beam emittance measurements.

This work describes physics-based models useful for investigating new photocathode gun designs. It suggests that further improvements in the performance of the photocathode gun can be made by simply reducing the beam size at the solenoid to minimize its aberrations. And non-uniform emission at the few percent level seeds space charge emittance roughly equal to the intrinsic emittance for 250 pC bunches.

In addition, expressions for the intrinsic emittance and QE are given which include effective mass effects of the pre-emission electron. This theory suggests using cathodes with large anisotropic effective masses is a possible approach to achieving ultra-low intrinsic emittance.

Recent advances in the space charge emittance compensation, symmetric RF fields, and beam dynamics of photocathode RF guns have enabled high gain free electron lasers to become productive 4$^{th}$ generation light sources. The next developments need to increase beam brightness by improving cathode performance of *QE* and intrinsic emittance, and by the elimination and correction of aberrations in the electron beam optics. Progress in these areas will produce ever more interesting physics and exciting new applications for high brightness electron beams.

**Acknowledgements**

The author wishes to express his appreciation for the comments and suggestions of Paul Emma, SLAC and John Schmerge, SSRL/SLAC and Ivan Bazarov, Cornell U. Paul's suggestions greatly improved the paper's clarity and directness. John got me started on trying to find simple relations for what appear to be complicated processes. I very much appreciate our common interest in electron injectors, and thank him for keeping me involved in the APEX/LCLS-II gun design and beam studies. And finally, I thank Ivan for his stimulating suggestions and probing comments.

**References**

[1] C. Gulliford, A. Bartnik, I. Bazarov, B. Dunham, and L. Cultrera, Demonstration of cathode emittance dominated high bunch charge beams in a DC gun-based photoinjector, Appl. Phys. Lett. 106, 094101 (2015).
[2] A. Bartnik, C. Gulliford, I. Bazarov, L. Cultera, and B. Dunham, Operational experience with nanocoulomb bunch charges in the Cornell photoinjector, Phys. Rev. ST Accel. Beams 18, 083401 (2015).
[3] The complete DC/RF photoinjector system is described by B. Dunham, Chapter 4 in *An Engineering Guide to Photoinjectors*, T. Rao and D. H. Dowell, editors. http://arxiv.org/abs/1403.7539
[4] L. Serafini and J. Rosenzweig, "Envelope analysis of intense relativistic quasi-laminar beams in RF photoinjectors: a theory of emittance compensation," *Phys. Rev. E*, vol. 55, pp. 7565-7590, June 1997.

[5] T. Schietinger, M. Pedrozzi, M. Aiba, V. Arsov, S. Bettoni *et al.,* "Commissioning experience and beam physics measurements at the SwissFEL Injector Test Facility," Phys. Rev. Accel. and Beams **19**, 100702 (2016).

[6] M. Ferrario, J. E. Clendenin, D. T. Palmer, J. B. Rosenzweig, and L. Serafini, Report No. SLAC-PUB8400, 2000 (unpublished).

[7] D. H. Dowell, I. Bazarov, B. Dunham, K. Harkay, C. Hernandez-Garcia, R. Legg, H. Padmore, T. Rao, J. Smedley, and W. Wan, "Cathode R&D for future light sources," Nucl. Instrum. Methods Phys. Res., Sect. **A 622**, 685 (2010).

[8] Proceedings of the Photocathode Physics for Photoinjectors, P3 Workshop held at Jefferson Lab in 2016, https://www.jlab.org/indico/event/124/ ; Proceedings for P3 Workshop held at LBNL, Berkeley, CA in 2014, https://sites.google.com/a/lbl.gov/photocathode-physics-for-photoinjectors/program.

[9] W.E. Spicer, "Photoemission, Photoconductivity, and Optical Absorption Studies of Alkali-Antimony Compounds," Phys. Rev. **112**, 114 (1958).

[10] Andrew Zangwell, *Physics at Surfaces* (Cambridge University Press, 1996) pp. 57-63.

[11] D.H. Dowell *et al*., PRST-AB **9**,63502(2006).

[12] D.H. Dowell and J.F. Schmerge, PRST-AB **12**,074201(2009) and references therein.

[13] "Excited-state thermionic emission in III-Antimonides: Low emittance ultrafast photocathodes," J.A. Berger, B.L. Rickman, T. Li, A.W. Nicholls, and W.A. Schroeder, Applied Physics Letters **101** (2012) 194103.

[14] C.P. Hauri *et al*., PRL **104**, 234802(2010).

[15] F. Zhou *et al*., PRST-AB, 5, 094203(2002).

[16] M. Reiser*, Theory and Design of Charged Particle Beams* (John Wiley & Sons, Inc., 1994) p. 470-479.

[17] A. Brachmann *et al*., Proceedings of FEL2009, pp. 463-465, Liverpool, UK.

[18] M. Reiser*, Theory and Design of Charged Particle Beams* (John Wiley & Sons, Inc, 1994) pp. 195-197.

[19] J. D. Lawson, J. Electron. Control **5**, 146 (1958)

[20] GPT: General Particle Tracer, Version 2.82, Pulsar Physics, http://www.pulsar.nl/gpt/

[21] K-J. Kim, NIM **A275**(1989)201-218.

[22] D.H. Dowell *et al.,* NIM A528(2004)316-320 and references therein.

[23] R. Akre *et al.,* PRST-AB **11**, 030703(2008).

[24] D.C. Cary, K.L. Brown and F. Rothhacker, "Third-Order TRANSPORT with MAD Input", SLAC-R-530, Fermilab-Pub-98-310, UC-414, page 161.

[25] D.H. Dowell, US Particle Accelerator School, June 14-18, 2010, Boston, MA.

[26] *An Engineering Handbook to Photoinjectors*, by T. Rao and D. H. Dowell, eds., published 2013
The pdf version of this book can be downloaded at http://arxiv.org/abs/1403.7539
The paperback version is available at the following Amazon.com link:
 http://www.amazon.com/s/ref=ntt_athr_dp_sr_1?_encoding=UTF8&field-author=Dr.%20Triveni%20Rao&search-alias=books&sort=relevancerank

[27] M. Reiser*, Theory and Design of Charged Particle Beams* (John Wiley & Sons, Inc., 1994) pp. 106-107.

[28] *General Particle Tracer, User Manual*, Version 2.82, p. 156.

[29] M. Born and E. Wolf, *Principles of Optics, Sixth Edition* (Pergamon Press, 1986) p. 217.

[30] D.H. Dowell *et al*., "The Development of the Linac Coherent Light Source RF Gun"

http://www-bd.fnal.gov/icfabd/Newsletter46.pdf & SLAC-Pub-13401 and references therein.

[31] D. H. Dowell, "Cancellation of RF Coupler-Induced Emittance Due to Astigmatism," pdf available at the Cornell Physics ArXiv, http://arxiv.org/abs/1503.09142 and SLAC Pubs Report-no: LCLS-II-TN-15-05.

[32] F. Zhou *et al*., "LCLS-II Injector Beamline Optimizations and RF Coupler Correction", Proc. of 2015 International FEL Conference.

[33] *General Particle Tracer, User Manual*, Version 2.82, pp. 138-139.